\documentclass[twocolumn,showpacs,eqsecnum,amsmath,amssymb,pra]{revtex4}

\usepackage{graphicx}
\usepackage{dcolumn}
\usepackage{bm}

\begin{document}


\title { 
Molecular terms, magnetic moments and optical transitions of 
molecular ions C$_{60}^{m\pm}$ 
}

\author {A.V. Nikolaev} 
 \altaffiliation[Also at:]{
Institute of Physical Chemistry of RAS, 
Leninskii prospect 31, 117915, Moscow, Russia 
} 

\author{K.H. Michel} 

\affiliation{%
Department of Physics, University of Antwerp, UIA, 2610, Antwerpen, Belgium
}%

\date{\today}

\begin{abstract} 
Starting from a multipole expansion of intra- molecular Coulomb
interactions, we present configuration interaction calculations
of the molecular energy terms of the hole configurations 
$(h_u^+)^m$, $m=2-5$, of C$_{60}^{m+}$ cations, of the electron
configurations $t_{1u}^n$, $n=2-4$, of the C$_{60}^{n-}$ anions,
and of the exciton configurations $(h_u^+ t_{1u}^-)$, $(h_u^+ t_{1g}^-)$ of
the neutral C$_{60}$ molecule.
The ground state of C$_{60}^{2-}$ is either $^3 T_{1g}$ or $^1A_g$,
depending on the energy separation between $t_{1g}$ and $t_{1u}$ levels. 
There are three close ($\sim$0.03 eV) low lying triplets
$^3 T_{1g}$, $^3 G_g$, $^3 T_{2g}$ for C$_{60}^{2+}$, and three
quartets $^4 T_{1u}$, $^4 G_u$, $^4 T_{2u}$ for C$_{60}^{3+}$,
which can be subjected to the Jahn-Teller effect.
The number of low lying nearly degenerate states in largest for $m=3$ holes.
We have calculated the magnetic moments of the hole and electron 
configurations and found that they are independent of molecular
orientation in respect to an external magnetic field.
The coupling of spin and orbital momenta differs from the atomic case.
We analyze the electronic dipolar transitions 
$(t_{1u})^2 \rightarrow t_{1u} t_{1g}$ and 
$(t_{1u})^3 \rightarrow (t_{1u})^2 t_{1g}$ for C$_{60}^{2-}$ and
C$_{60}^{3-}$. 
Three optical absorption lines 
($^3 T_{1g} \rightarrow {}^3 H_u, {}^3 T_{1u}, {}^3 A_u$)
are found for the ground level of C$_{60}^{2-}$ and only one line 
($^4 A_u \rightarrow {}^4 T_{1g}$) for the ground state of C$_{60}^{3-}$.
We compare our results with the experimental data for C$_{60}^{n-}$
in solutions and with earlier theoretical studies. 
\end{abstract} 
 

\maketitle

\section {Introduction} 
\label{sec:int}

The physics and chemistry of fullerenes is full of surprises.
After almost ten years of intensive theoretical and
experimental work, unexpected discoveries of 
ferromagnetic polymerized C$_{60}$ \cite{Mak} as well as
superconductivity
in hole doped pristine C$_{60}$ \cite{Bat1} and in lattice-expanded C$_{60}$
\cite{Bat2} raise new questions about the electronic structure
of the C$_{60}$ molecule and its molecular ions. 
Using the field-effect doping techniques, it has been shown
that high transition temperatures are achieved for the case
of three holes per the C$_{60}$ molecule \cite{Bat1}, when many
electron effects come into play. 
Motivated by these new experimental findings, 
here we present a calculation of the electronic structure of few holes 
of C$_{60}^{m+}$.
Our second goal is to study the molecular term picture of 
the C$_{60}^{n-}$ anion
($n=2,3,4$), which behaves as a quasi-element in a vast majority
of ionic compounds \cite{Tan}. 
When C$_{60}^{m\pm}$ units are in a solid, 
additional inter-molecular interactions superimposed 
on the intra-molecular ones should be taken into account 
and a realistic theory of solids should treat both kinds 
of interactions on equal footing.
Therefore, the consideration of intra- molecular correlations
should be a necessary ingredient of a many electron theory aiming to
describe superconductivity, magnetic properties 
or a metal-insulator transition \cite{Tan,Gun1,Dre,For,Kuz}. 
The problem is
also a challenge from experimental point of view \cite{Reed,Boyle}.
In the present paper we will give quantitative results on the many electron
terms and magnetic moments of C$_{60}^{m+}$ and C$_{60}^{n-}$.

While the one-electronic structure of the neutral C$_{60}$ molecule 
is known for many years \cite{Had} (in fact, even before the actual
discovery of the Buckminsterfullerene \cite{Boc})
the case of two or three electrons (holes)
on the degenerate $t_{1u}$ or $h_u$ shell requires a special treatment.
Then the two ($n=2$) or three ($n=3$) electrons (holes) are equivalent 
and should be treated on equal footing.
Such kind of situation is very typical for atoms with open electron shells.
When an atom has two or more valence electrons on orbitally degenerate 
states (like $p$, $d$ or $f$) its energy spectrum
can be very complex reflecting the electronic degrees of freedom.
There exist empirical observations known as Hund's rules, 
which prescribe the occupation of the orbitals, but those 
are just consequences of the atomic theory of many electron states 
formulated by Condon and Shortley long ago \cite{CS}.
The real driving force behind the term splitting is the Coulomb repulsion
of the valence electrons.
While in atoms electronic energy levels are split due to the multipolar Coulomb
intra-atomic interactions, in the case of C$_{60}$ we deal with
the intra-molecular ones. 

Here we present an approach which was inspired by the theory of
many electron atomic states \cite{CS}.
It is also an extension of our original method of multipole expansion
for electronic states \cite{Nik,NM1}.
It is worth mentioning that
our treatment should not be confused with an ordinary single-determinant
Hartree-Fock calculation which does not
take into account the intra-molecular correlations and the 
molecular term structure is overlooked. 
We will see (Sec.\ III) that in our approach
each basis function represents a Slater determinant, and the
solution is found as their linear combination.
As such, the method corresponds to a many determinant treatment
or configuration interaction (CI). Therefore, the approach is a
genuine many electron one as long as we limit ourselves to the
relevant orbital space ($h_u$, $t_{1u}$ or $t_{1g}$). 
To our knowledge, in the literature there are only two calculations
concerning the electronic structure of negatively charged
C$_{60}^{n-}$ molecular ions, reported by Negri {\it et al.} \cite{Neg}, 
and by Saito {\it et al.} \cite{Sai}. 
In the latter work, however, the assignment of excitations with
molecular terms is done only for the neutral molecule ($n=0$).

In the present paper we assume that the molecule has the
icosahedral symmetry. 
If one wants to consider
a distortion of C$_{60}^{m+}$ or C$_{60}^{n-}$ \cite{Ceul,Tos2,Dun2,Bri,Moa}, 
the computed energy levels and their eigenvectors can be used as
a starting point for the description of the Jahn-Teller effect
in these systems. Indeed, the electron-phonon (or vibronic)
coupling occurs if $[\Gamma_{el}]^2$ contains $\Gamma_{vib}$ \cite{Lan,Ber}.
Here $\Gamma_{el}$ is the symmetry of the electronic molecular
term under consideration, while $\Gamma_{vib}$ is the symmetry of a
vibrational normal mode. It is evident then that for a meaningful
analysis of the vibrational coupling and the resultant
Jahn-Teller distortion, one has to know the symmetry of the
corresponding electronic terms, that is the issue of the
present study. In addition, now it has been realized
that the Jahn-Teller effect for C$_{60}^{-}$ is rather weak,
and the situation is probably best described as dynamic \cite{Dun1},
where the molecule on average retains its icosahedral symmetry.
The vibronic coupling for C$_{60}^{m+}$ is not so well investigated,
but estimated as $\sim0.1-0.2$ eV. We will see that the energy span
of the hole configurations a few times exceeds this value.
Therefore, if there is a static Jahn-Teller effect for C$_{60}^{m+}$,
it can be treated in the limit of weak or intermediate coupling,
leading to a more complex picture where the molecular terms
are subjected to further splitting \cite{Bra,Ber}.

Since the C$_{60}$ molecule reveals nontrivial degeneracies
and peculiarities, its term structure has been studied by group-theoretical
methods \cite{Pla1,Oli,Lo,Jud,Pla2}.
The analysis there was
focused on the symmetry and number of terms for $g^N\,(N=1-7)$ \cite{Pla1} and 
$h^N\,(N=1-9)$ \cite{Oli,Lo,Jud,Pla2} configurations, but by itself
it can not give a quantitative picture of the splittings.
Where our approach overlaps with the group-theoretical one, our
findings are in agreement with the latter.

The paper comprises the following sections.
First (Sec.\ II), 
we introduce the angular dependencies of $h_u$, $t_{1u}$ and $t_{1g}$
functions.
Next (Sec.\ III), we describe our method of treating
multipole Coulomb correlations for many electrons. 
Then we apply it to hole- and excitonic configurations of
the C$_{60}$ molecule, Sec.\ IV.
In Sec.\ V we give the resultant energy spectra
and compare our calculations with those of 
Negri {\it et al.} \cite{Neg} for C$_{60}^{2-}$.
In Sec.\ VI we compute magnetic moments, 
in Sec.\ VII optical lines and line strengths for
the electron dipolar transitions of C$_{60}^{2-}$ and C$_{60}^{3-}$.
Finally, we give our conclusions in Sec.~VIII.

\section {Angular dependence of molecular orbitals} 
\label{sec:orb} 

The neutral C$_{60}$ molecule has the highest molecular point group
($I_h$). Expanding its density in multipole series one
finds that nontrivial angular dependencies are given by the
symmetry adapted functions (SAFs) \cite{Bra} of $A_{1g}$ symmetry with 
the lowest components characterized by $l=6$, $l=10$ 
and $l=12$ \cite{Coh,Mic1}. That makes the C$_{60}$ fullerene the most
spherical molecule among the others.
Owing to such unique symmetry, the parentage of $\pi$-molecular
orbitals in spherical harmonics can
be clearly traced \cite{Had}. For lowest occupied $\pi$-levels of the
neutral molecule we have $a_g\, (l=0)$, $t_{1u}\, (l=1)$, $h_g\,(l=2)$,
$t_{2u}\, (l=3)$, $g_u\, (l=3)$, $g_g\, (l=4)$, $h_g\, (l=4)$ and
$h_u\, (l=5)$. The corresponding molecular states accommodate 60 
electrons and the seven electron shells are completely filled.
Then the generalized Uns\"old theorem \cite{Tin} ensures that the resulting
electron density of the $\pi$-states has full (or unit) icosahedral
symmetry $A_{1g}$. (Here and below we use capital letters for
the irreducible representations (irreps) of density and small letters for
the irreps of electron wave functions.)

Due to the direct correspondence between these $\pi$-shells and the
molecular orbital index $l$, one can immediately find out an
orbital part for a given $\pi$-state in the same way as we know
the orbital parts of $s$, $p$, $d$ and $f$ electrons in an atom.
Therefore, the type of an irreducible representation ($a$, $t_1$, $t_2$, $g$
and $h$) and the orbital index $l$ uniquely determine the angular dependence of
the molecular orbitals. The angular functions are called spherical harmonics
adapted for the icosahedral group $I_h$. Such symmetry adapted
functions (SAFs) were tabulated by Cohan in Ref.\ \onlinecite{Coh}
for all $l \le 14$. (Unfortunately, Cohan worked with unnormalized
spherical harmonics so that it requires some efforts to express SAFs
in conventional spherical harmonics $Y_l^m$.)
In particular for the highest occupied molecular orbital (HOMO) $h_u\, (l=5)$
one has
\begin{subequations}
\begin{eqnarray}
  & & \psi_1(h_u)=Y_5^{5,s} , \label{1.1a} \\
  & & \psi_2(h_u)=\sqrt{\frac{7}{10}} Y_5^{1,c}+\sqrt{\frac{3}{10}} Y_5^{4,c} , \\
  & & \psi_3(h_u)=\sqrt{\frac{7}{10}} Y_5^{1,s}-\sqrt{\frac{3}{10}} Y_5^{4,s} , \\
  & & \psi_4(h_u)=\sqrt{\frac{2}{5}} Y_5^{2,c}+\sqrt{\frac{3}{5}} Y_5^{3,c} , \\
  & & \psi_5(h_u)=\sqrt{\frac{2}{5}} Y_5^{2,s}-\sqrt{\frac{3}{5}} Y_5^{3,s} .
\label{1.1e}
\end{eqnarray}
\end{subequations}
Here the normalized real spherical harmonics are defined with the phase convention
of Ref.\ \onlinecite{Bra} and the orientation of the C$_{60}$ molecule
corresponds to the choice of the $z$ axis as one of 12 fivefold axes
and the $y$ axis as one of the twofold axes perpendicular to $z$ \cite{Coh}.
We call this position of C$_{60}$ the orientation of Cohan.
In order to transform the molecule to the
standard orientation \cite{Dav} where molecular twofold axes lie along the
Cartesian $x$, $y$ and $z$ direction one has to rotate the molecule anticlockwise 
about the $y$ axis by an angle
$\beta=arccos(2/\sqrt{10+2\sqrt{5}}) \approx 58.28^0$ \cite{Nik}. 
Each of the $h_u$ orbital function then is expressed in terms of $Y_5^{\tau}$,
where $\tau$ stands for $m=0$ or $(m,c)$, $(m,s)$ of real spherical harmonics
(Appendix B). In the following we will work with 
the molecule in the orientation of Cohan. (Of course, the results are independent of
the choice of the molecular orientation.)

The lowest unoccupied molecular $t_{1u}$ orbital (LUMO) corresponds
to $l=5$ and has the following three angular components in Cohan's 
orientation of C$_{60}$: 
\begin{subequations}
\begin{eqnarray}
  & & \psi_1(t_{1u})=\frac{6}{\sqrt{50}}Y_5^0+\sqrt{\frac{7}{25}}Y_5^{5,c} , 
  \label{1.2a} \\
  & & \psi_2(t_{1u})=\sqrt{\frac{3}{10}} Y_5^{1,c}-\sqrt{\frac{7}{10}} Y_5^{4,c} , \\
  & & \psi_3(t_{1u})=\sqrt{\frac{3}{10}} Y_5^{1,s}+\sqrt{\frac{7}{10}} Y_5^{4,s} .
\label{1.2c}
\end{eqnarray}
\end{subequations}
The angular parts of $t_{1u}$ LUMO have been derived before in Refs.~\onlinecite{Tos}
and \onlinecite{Nik}. In the standard orientation of C$_{60}$ they are
given by Table~I of Ref. \onlinecite{Nik}.
The LUMO-HOMO energy gap is about 2.7 eV \cite{Reed,Dre,Yang}.

Finally, at an energy about 1.15 eV \cite{Reed,Kato} above LUMO 
one finds the molecular $t_{1g}$
level with $l=6$ (LUMO+1). In the orientation of Cohan the
angular parts are given by
\begin{subequations}
\begin{eqnarray}
  & & \psi_1(t_{1g})= Y_6^{5,s} , \label{1.3a} \\
  & & \psi_2(t_{1g})=\sqrt{\frac{11}{2}} \frac{\sqrt{3}}{5} Y_6^{1,c}-
  \sqrt{\frac{11}{2}}\frac{1}{5} Y_6^{4,c}+\frac{\sqrt{3}}{5} Y_6^{6,c} , 
        \nonumber \\
  & & \label{1.3b} \\
  & & \psi_3(t_{1g})=\sqrt{\frac{11}{2}} \frac{\sqrt{3}}{5} Y_6^{1,s}+
  \sqrt{\frac{11}{2}}\frac{1}{5} Y_6^{4,s}+\frac{\sqrt{3}}{5} Y_6^{6,s} . 
   \nonumber \\
  & & \label{1.3c}
\end{eqnarray}
\end{subequations}

\section {Method of calculation} 
\label{sec:met} 

Our method of multipole expansion of the Coulomb interaction has been
reported before \cite{Nik,NM1}.
Here we extend it and apply to the case of the C$_{60}$ molecule with
the icosahedral symmetry.
In the following we consider in detail the case of two and three $t_{1u}$ electrons.
Starting with a pair of electrons we will give a special attention 
to the procedure of adding one extra $t_{1u}$ electron to the pair. 
In the same way one can add a fourth electron
to the group of three electrons and etc. Therefore,
our main goal of treating $n$ electrons 
can be reached by adding one electron after another.

Since the estimated one-electron spin-orbit coupling is negligible
($\sim 0.16$ cm$^{-1}$) \cite{Tos},
 we are working in the ``$LS$ (Russell-Saunders)"
molecular approximation. (The spin-orbit coupling is a single particle
operator and in principle can be included in the calculation \cite{NM1}.)
We start with a pair of $t_{1u}$ electrons and label the two-electron basis 
ket-vectors by a single index
$I$ which incorporates a pair of one-electron indices $(i_1,i_2)$,
\begin{eqnarray}
   | I \rangle =  | i_1; \, i_2 \rangle .
\label{2.1} 
\end{eqnarray}
The indices $i=(k,s_z)$ stand for the $t_{1u}$ orbitals ($k=1,2,3$)
and the spin projection quantum number. The corresponding basis wave
functions are 
\begin{eqnarray}
  \langle \vec{r}, \vec{r}\,'  | I \rangle = 
  \langle \vec{r}\, | i_1 \rangle \cdot
  \langle \vec{r}\,' | i_2 \rangle ,
\label{2.2} 
\end{eqnarray}
where $\langle \vec{r}\, | i \rangle = {\cal R}(r) \langle \hat{n} | i \rangle$.
Here ${\cal R}$ is the radial component of the $t_{1u}$ 
molecular orbitals (MO),
$\hat{n}$ stands for polar angles $\Omega=(\Theta,\phi)$.
There are six orientational $t_{1u}$ vectors (or spin-orbitals)
$\langle \hat{n} | i \rangle$ ($i.e.$ $i=1-6$),
\begin{eqnarray}
  \langle \hat{n} | i \rangle=\psi_k(\hat{n})\, u_s(s_z) . 
 \label{2.4}  
\end{eqnarray}
Here $\psi_k$ are the three $t_{1u}$ MOs as given by Eqs.~(\ref{1.2a}-c)
for the Cohan's orientation of C$_{60}$, 
$u_s$ is the spin function ($s=\pm$) for the spin projections
$s_z=\pm1/2$ on the $z$-axis.

The order of indices in (\ref{2.1}) and (\ref{2.2}) is important 
if we associate the first electron with the state $i_1$ while
the second with the state $i_2$. 
From the dynamical
equivalence of the electrons we can permute the spin-orbitals
of the state $| i_2;\, i_1 \rangle$ 
to the standard order, Eq.\ (\ref{2.1}), by using
\begin{eqnarray}
   | i_2; \, i_1 \rangle  = -| i_1; \, i_2 \rangle  ,
\label{2.5} 
\end{eqnarray}
since it requires the interchange of the two electrons.
In order to describe the same quantum state we will use
the basis vectors (\ref{2.1}) where $i_1>i_2$ and apply Eq.\ (\ref{2.5}) when 
needed.
(Alternatively, one can use the standard
procedure of antisymmetrization of the basis vectors (\ref{2.1}).) 
Thus, our basis (\ref{2.1}) consists of 
$(6\times 5)/2=$15 different vectors 
$| I \rangle$. 

In the following we will study the intra- molecular correlations
of electrons within a formalism based on a multipole expansion
of the Coulomb potential between two electrons (charge $e=-1$),
\begin{eqnarray}
 V(\vec{r},\vec{r}\,')=\frac{1}{|\vec{r} - \vec{r}\,' |} .
\label{2.6} 
\end{eqnarray}
The multipole expansion in terms of real spherical harmonics
$Y_l^0$, $Y_l^{m,c}$ and $Y_l^{m,s}$ 
(we use the phase convention and the definitions 
of Ref.\ \cite{Bra}) reads:
\begin{eqnarray}
  V(\vec{r},\vec{r}\,')=
  \sum_{l,\tau}  v_{l}(r,r')\,
  Y_l^{\tau}(\hat{n})\, Y_l^{\tau}(\hat{n}'),  
\label{2.7} 
\end{eqnarray}
where $\tau$ stands for $m=0$, $(m,c)$ or $(m,s)$ of the real spherical
harmonics and
\begin{eqnarray}
  v_{l}(r,r')\,
  = \left( {\frac {r^l_< }{r^{(l+1)}_> }} \right)
  {\frac {4\pi}{2l+1}}  , 
\label{2.8} 
\end{eqnarray}
with $r_>=max(r,r')$, $r_<=min(r,r')$.

The direct matrix elements for the intra-molecular Coulomb
interactions are obtained if we consider
the $i_1 \rightarrow j_1$ transitions for the first electron
and the $i_2 \rightarrow j_2$ transitions for the second
(we recall that $i_1>i_2$ and $j_1>j_2$). We label this
two-electron transition by the index $a_2=1$.
Starting from Eq.~(\ref{2.7}) we obtain
\begin{eqnarray}
 \langle I  | V(\vec{r},\vec{r}\,') | J \rangle^{Coul} =
  \sum_{l,\tau} 
  v_{l} \, c_{l,\tau}(i_1 j_1)\, c_{l,\tau}(i_2 j_2), 
\label{2.9} 
\end{eqnarray}
where
\begin{eqnarray}
  v_{l} =
  \int \! dr\, r^2 \int \! dr'\, {r'}^2\,  
  {\cal R}^2(r)\, {\cal R}^2(r')\,
  v_{l}(r,r')  
\label{2.10} 
\end{eqnarray}
accounts for the average radial dependence.
The transition matrix elements $c_{l,\tau}$ are defined by
\begin{eqnarray}
  c_{l,\tau}(i j) =  
  \int \! d\Omega \, \langle i |\hat{n}\rangle \, 
  Y_l^{\tau} (\hat{n}) \, \langle \hat{n}|j \rangle .
\label{2.11} 
\end{eqnarray}

The other possibility is to consider the transitions
$i_1 \rightarrow j_2$ for the first electron and the
transitions $i_2 \rightarrow j_1$ for the second.
We label it by the index $a_2=2$.
This gives the exchange interaction and then we 
use (\ref{2.5}) to return to the standard order
of the spin-orbitals. We find
\begin{eqnarray}
 \langle I | V(\vec{r},\vec{r}\,') | J \rangle^{exch} =
  -\sum_{l,\tau}  v_{l} \,
  c_{l,\tau}(i_1 j_2)\, c_{l,\tau}(i_2 j_1) , \quad
\label{2.12} 
\end{eqnarray}
where $v_l$ again is given by Eq.\ (\ref{2.10}) and
the coefficients $c_{l,\tau}$ by Eq.\ (\ref{2.11}).
We observe that in the basis with the real $t_{1u}$ orbitals,
and with the real spherical harmonics $Y_l^{\tau}$
the coefficients $c_{l,\tau}$ are real.

We start with the spherically symmetric term ($l=0$)
corresponding to the trivial function $Y_0^0=1/\sqrt{4\pi}$.
The coefficients $c_{l,\tau}$ in (\ref{2.11}) become diagonal,
$c_{l=0}(ij)=1/\sqrt{4\pi} \delta_{ij}$.
In considering the other contributions (with $l>0$)
we take advantage of the selection rules imposed by
the coefficients $c_{l,\tau}$, Eq.\ (\ref{2.11}).
First of all, we notice that the coefficients $c_{l,\tau}$
are diagonal in terms of spin components $u_s$.
Secondly, the odd values of $l$ are excluded due to the
parity of the integrand in (\ref{2.11}), and from the theory
of addition of angular momenta we know that $l_{max}=10$. 
From these observations we conclude that $l=0$, 2, 4, 6, 8 and 10.
Earlier the coefficients $c_{l,\tau}$ have been used 
for the description of the crystal field of C$_{60}^{-}$ in Ref.\ \cite{Nik}.
By adding (\ref{2.12}) to (\ref{2.9}) we obtain
\begin{eqnarray}
  & & \langle I | V(\vec{r},\vec{r}\,') | J \rangle  = 
  U_0\, \delta(I,J) +v_2  c_2(I | J ) + v_4  c_4(I|J) \nonumber \\ 
  & & 
  + v_6  c_6(I|J) +v_8  c_8(I|J) + v_{10}  c_{10}(I|J) ,
\label{2.12'} 
\end{eqnarray}
where $U_0=v_0/4\pi$ is the Hubbard repulsion, $\delta$ the Kronecker symbol, and
\begin{eqnarray}
 c_l(I | J ) & = &  \sum_{\tau} [ 
 c_{(l,\tau)}(i_1 j_1)\,c_{(l,\tau)}(i_2 j_2)   \nonumber \\
 & & -c_{(l,\tau)}(i_1 j_2)\,c_{(l,\tau)}(i_2 j_1) ] .
\label{2.14} 
\end{eqnarray}
We have studied the secular problem for the $15\times15$ matrix
of intra-molecular interactions and obtained 15 energy levels 
$E_p[t_{1u}^2]$ ($p=1-15$),  
\begin{eqnarray}
  & & E_p[t_{1u}^2]  =  U_0 +v_2  \lambda_2(p) + v_4  
   \lambda_4(p) \nonumber \\ 
  & & + v_6  \lambda_6(p) +v_8  \lambda_8(p) +
         v_{10}  \lambda_{10}(p) , \quad
\label{2.15} 
\end{eqnarray}
where $\lambda_l(p)$ 
are numerical constants (called ``integral" or ``molecular invariants"
in Refs.\ \cite{Pla1,Oli,Lo,Pla2}). They are quoted in Table I.
%
\begin{table} 
\caption{ 
 Coefficients $\lambda'_l(p)=\lambda_l(p) \times 10^3$ for
 $(t_{1u})^2$ ($p=1-15$) and $(t_{1u})^3$ ($p'=1-20$).
\label{table1}     } 
\begin{ruledtabular}
 \begin{tabular}{c c c r r r r r } 
  & $p$ & deg. & $l=2$ & 4 & 6 & 8 & 10 \\ 
\hline 
           & 1-9   & (9) & -2.943 & -21.190 & 21.817 & -39.706 & -37.556  \\ 
$\lambda'_l$& 10-14 & (5) &  0.589 &   4.238 & 23.596 &   7.941 &  11.641  \\
           & 15    & (1) &  5.886 &  42.379 & 26.265 &  79.413 &  85.436  \\
\hline
          & 1-4   & (4) & -8.829 & -63.568 & 65.450 & -119.119 & -112.667 \\
$\lambda'_l$  & 5-14  & (10)& -3.532 & -25.427 & 68.119 &  -47.648 &  -38.872 \\
          & 15-20 & (6) &  0.0   &   0.0   & 69.898 &    0.0   &   10.324
 
 \end{tabular} 
\end{ruledtabular}
\end{table} 
The 15 levels of (\ref{2.15}) form three distinct terms, $i.e.$
a 9-fold degenerate $\{ t_{1u}^2;1\}$, a 5-fold degenerate $\{ t_{1u}^2;2\}$
and a single-level term $\{ t_{1u}^2;3\}$.
The symmetry between the four electrons and the two holes within
the $t_{1u}$ LUMO configuration implies 
$\lambda_l(t_{1u}^4;\,p)=\lambda_l(t_{1u}^2;\,p)$, $p=1-15$.
Therefore, Eq.~(\ref{2.15}) holds also for the case of four $t_{1u}$ 
electrons if we write $6U_0$ in place of $U_0$.
In order to study the splitting quantitatively, one has to calculate the
radial integrals $v_l$, Eq.~(\ref{2.10}). We leave this calculation
for next section and now move on to the case of three $t_{1u}$ electrons.

For the $t_{1u}^3$ there are $6\times5\times4/3!=20$ 
basis ket-vectors $| I' \rangle =| i_1 i_2 i_3 \rangle$,
where $i_1 > i_2 > i_3$.
Now for the Coulomb interaction we have a sum of three two-body terms, 
\begin{eqnarray}
 V^{(3)}=V(\vec{r}_1,\vec{r}_2)+V(\vec{r}_1,\vec{r}_3)+V(\vec{r}_2,\vec{r}_3) ,
\label{2.14n} 
\end{eqnarray}
where each $V(\vec{r}_a,\vec{r}_b)$ is given by the multipole expansion (\ref{2.7}).
In considering a matrix element $\langle I' |V^{(3)}| J' \rangle$ we
have many subcases which we also call transitions. 
We sort them out as shown in Fig.~1. 
%
\begin{figure} 
\resizebox{0.46\textwidth}{!}
{ 
 \includegraphics{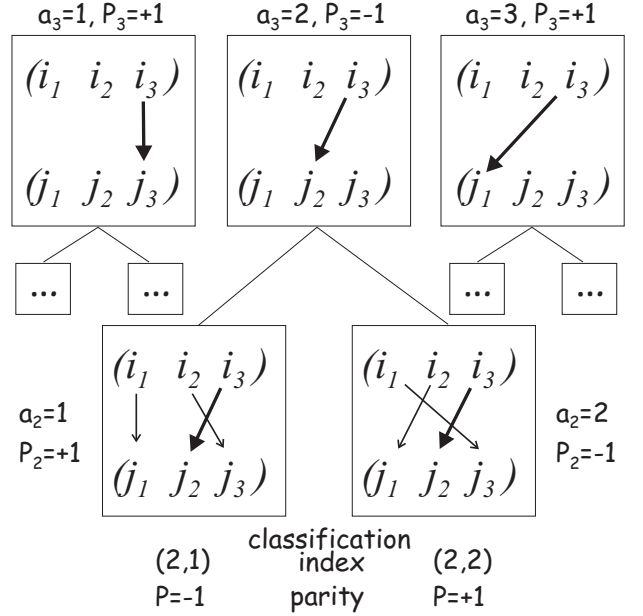} 
} 
\vspace{2mm}
\caption{
Diagram for the calculation of a matrix element 
$\langle i_1 i_2 i_3 | V^{(3)} | j_1 j_2 j_3 \rangle$ by expressing it 
in terms of different one-electron transitions $i \rightarrow j$. 
See text for details.
} 
\label{fig1} 
\end{figure} 
%
It is important to notice that for 
the third electron there are only three possibilities, i.e.
$i_3 \rightarrow j_3$, $i_3 \rightarrow j_2$ and $i_3 \rightarrow j_1$,
which are labeled by $a_3=1,2$ and 3, respectively.
From the anticommutation relations we find that the parity ($P_3$) of the transitions 
are $+1$, $-1$ and $+1$. 
Each case leaves possibilities for further
transitions of the remaining two $t_{1u}$
electrons. We denote their final states as $j'_1$ and $j'_2$.
(For $a_3=1$ these indices are $j_1$ and $j_2$;
for $a_3=2$ $j_1$ and $j_3$, Fig.~1; 
and for $a_3=3$ $j_2$ and $j_3$.)
However, the two-electron transitions 
have been sorted out before when we considered the $t_{1u}^2$ configuration. 
In that case there are only two options, 
($a_2=1$) $i_1 \rightarrow j'_1$, $i_2 \rightarrow j'_2$,
and ($a_2=2$) $i_1 \rightarrow j'_2$, $i_2 \rightarrow j'_1$. 
The first has parity $P_2=+1$, the second $P_2=-1$. 

Therefore, we can classify the transitions by two indices $(a_3,a_2)$ 
where $a_3=1-3$ and $a_2=1,2$.
For a general case of $n$ electrons it is $(a_n,a_{n-1},a_{n-2},...,a_2)$, 
where $a_k=1,2,...,k$.
The parity of an $(a_3,a_2)$ transition is $P=P_3(a_3)\, P_2(a_2)$, 
for the general case
$P=P_n(a_n) P_{n-1}(a_{n-1})\cdot ... \cdot P_2(a_2)$.
The total number of transitions of three $t_{1u}$ electrons is $3 \times 2=6$,
in general it is $n!$. For each of the subcases we calculate the matrix element
of $V^{(3)}$, Eq.~(\ref{2.14n}),
\begin{eqnarray}
& &  \langle I'  | V^{(3)} | J' \rangle^{(a_3,a_2)} = P_3(a_3)\,P_2(a_2)
 \nonumber \\
 & &  \times  \sum_{l,\tau} 
  v_{l} \, c_{l,\tau}(i_1 j'_1)\, c_{l,\tau}(i_2 j'_2)\, \delta(i_3,j_{a_3}) 
  +\,p.i. \quad  
\label{2.15n} 
\end{eqnarray}
Here $p.i.$ stands for the other pair Coulomb interactions
$V(\vec{r}_a,\vec{r}_b)$. For three particles there are three terms, Eq.\
(\ref{2.14n}).  
The first term is given in Eq.\ (\ref{2.15n}), two others are found by
replacing
$c_{l,\tau}(i_1 j'_1)$ $c_{l,\tau}(i_2 j'_2)$ $\delta(i_3,j_{a_3})$ 
 with
$c_{l,\tau}(i_1 j'_1)$ $c_{l,\tau}(i_3 j_{a_3})$ $\delta(i_2,j'_2 )$
and with $c_{l,\tau}(i_2 j'_2)$ $c_{l,\tau}(i_3 j_{a_3})$ $\delta(i_1,j'_1)$.
(The parity of the transitions of course remains the same.)
For the general case of $n$ electrons one has $( {}_2^{n} )$ different pairs 
for each $n$-electron transition. 
(Here $({}^n_2)$ is a binomial coefficient.)
Finally, the matrix element $\langle I' |V^{(3)}| J' \rangle$ is found as
\begin{eqnarray}
 \langle I' |V^{(3)}| J' \rangle = \sum_{a_3=1}^3 \sum_{a_2=1}^2 
 \langle I'  | V^{(3)} | J' \rangle^{(a_3,a_2)} .
\label{2.16n} 
\end{eqnarray}
Having found the matrix elements, 
we solve a $20 \times 20$ secular problem for $t_{1u}^3$.
The energy levels are given by
\begin{eqnarray}
  E_{p'}[t_{1u}^3]  &= & \left( {}^{n=3}_{\;\;\;2} \right) 
  U_0 + \sum_{l=2}^{10} v_l  \lambda_l(p') ,
\label{2.20} 
\end{eqnarray}
where only even values of $l$ occur, $p'=1-20$, 
and $\lambda_l$ are quoted in Table I. 
There are three terms,
a 4-fold degenerate $\{ t_{1u}^3;1\}$ (ground state), a 
10-fold degenerate $\{ t_{1u}^3;2\}$ 
and a 6-fold degenerate $\{ t_{1u}^3;3\}$, see Table~I.

The classification scheme described above is very useful for
handling a single-particle interaction ${\cal A}$. 
In particular, the electron coupling to an
external magnetic field, the spin-orbit interaction,
crystal electric field effects and the electron dipolar
operator fall in this class.
The main difference from the Coulomb case is that now the interaction
occurs to a single electron (represented by an arrow in Fig.~1) while
the rest (the other $(n-1)$ electrons, or arrows in Fig.~1) 
produce $(n-1)$ Kronecker factors.
For example, for the three $t_{1u}$ electrons we have
\begin{eqnarray}
 {\cal A}=A_1 + A_2 + A_3 ,
\label{2.17n} 
\end{eqnarray}
where $A_k$ refers to a single electron.
For each three-electron subcase $(a_3,a_2)$, Fig.~1, we obtain 
\begin{eqnarray}
& &  \langle I'  | {\cal A} | J' \rangle^{(a_3,a_2)} = P_3(a_3)\,P_2(a_2)
 \nonumber \\
 & &  \times  \sum_{l,\tau} 
    \langle i_1 |A_1| j'_1\rangle \, \delta(i_2 j'_2)\, \delta(i_3,j_{a_3}) 
    +\,c.p.\, . \quad  
\label{2.18n} 
\end{eqnarray}
$c.p.$ stands for the two other terms obtained from the first by
 two cyclic permutations, $i.e.$ when
$i_1 \rightarrow i_2 \rightarrow i_3$ and 
$j'_1 \rightarrow j'_2 \rightarrow j_{a_3}$.
For the general $n$-electron case there will be $n$ such terms for each 
matrix element $\langle I'  | {\cal A} | J' \rangle^{(a_n,a_{n-1},...,a_2)}$.

Finally, we would like to mention that an operator $R$ (rotation or inversion)
of the icosahedral group $I_h$ acts on all electrons simultaneously
and therefore can be written
as $R=R_n\, R_{n-1} \cdot ... \cdot R_1$, where $R_k$ stands for the
corresponding one-particle operator.
The classification scheme again is useful for determining the matrix elements
of the transformation in the many electron space. For example,
for three $t_{1u}$ electrons we obtain
\begin{eqnarray}
  \langle I | R | J \rangle = \sum_{a_3=1}^3 \sum_{a_2=1}^2
  \langle I | R | J \rangle^{(a_3,a_2)} ,
\label{2.19n} 
\end{eqnarray}
where
\begin{equation}
  \langle I | R | J \rangle^{(a_3,a_2)} = P^{(a_3,a_2)}
  \langle i_1 |R_1| j'_1 \rangle \langle i_2 |R_2| j'_2 \rangle
  \langle i_3 |R_3| j_{a_3} \rangle . 
 \nonumber 
\end{equation}
Expressions of that type were used to calculate the characters of
molecular terms and to identify their symmetry.
The orbital part of the many electron wave function
transforms as an irrep of $I_h$, while the spin function as 
a single ($t_{1u}^2$) or double valued ($t_{1u}^3$) representation of $SO(3)$.
Therefore, we classify the molecular terms \cite{Pla1,Oli,Lo,Jud} 
by the symbol $^{2{\cal S}+1} \Gamma$, 
where $2{\cal S}+1$ is the spin multiplicity and
$\Gamma$ is an irrep of $I_h$.
Thus, the molecular terms are ${}^3 T_{1g}$, ${}^1 H_{g}$, ${}^1 A_{g}$
for $t_{1u}^2$, and ${}^4 A_{u}$, ${}^2 H_{u}$, ${}^2 T_{1u}$ for $t_{1u}^3$,
Table \ref{table1}.

\section {Application to other configurations} 
\label{sec:app} 

Below we apply our method for the hole configurations $(h_u^+)^m$ 
of C$_{60}^{m+}$
and excitonic configurations $h_u^+ t_{1u}^-$, $h_u^+ t_{1g}^-$ of the
neutral molecule.
We also consider configurations $(t_{1u} t_{1g})$ and $(t_{1u})^2 t_{1g}$,
because they are important for calculations of
electronic dipolar transitions $(t_{1u})^2 \rightarrow t_{1u} t_{1g}$
and $(t_{1u})^3 \rightarrow (t_{1u})^2 t_{1g}$ of the anions C$_{60}^{2-}$
and C$_{60}^{3-}$.

\subsection{Hole configurations $(h_u^+)^m$ ($m=2-5$) }

The treatment of a $m$-hole configuration $(h_u^+)^m$ formally
coincides \cite{CS} with the analogous electronic case $(h_u)^m$, i.e. 
the Coulomb repulsion between holes and between electrons is the same.
The only difference concerns the spherically symmetric term
(Hubbard $U_0$). Below we count the energy from the level of an empty
$h_u$ shell. In constructing the basis functions, one should take
into account that the one-hole index $i_{hu}^+$ changes from one
to ten, where the five-fold degeneracy is due to the orbital
freedom, Eqs.~(\ref{1.1a}-e), and the two-fold degeneracy due to its spin.  
As a result we obtain the term energies
\begin{eqnarray}
  E_p[h_u^{+m}]  = \left( {}^{10-m}_{\;\;\;2} \right) U_0 + 
  \sum_{l=2}^{10} v_l  \mu_l^{(m)}(p) ,
\label{2.1sub} 
\end{eqnarray}
where only even values of $l$ occur.
The coefficients $\mu_l^{(m)}(p)$ 
are again molecular invariants. We quote them in Tables \ref{table5} and 
\ref{table6} for two and three holes, configurations 
$(h_u^+)^2$ and $(h_u^+)^3$.
%
\begin{table} 
\caption{ 
 Coefficients $\mu_l(p) \times 10^3$ 
 for $(h_u^+)^2$. The coefficients
 marked by $(*)$ are not unique depending on integrals $v_l$, 
 see text for details.
\label{table5}     } 
\begin{ruledtabular}
 \begin{tabular}{l c c r r r r r } 
  &  & deg. & $l=2$ & 4 & 6 & 8 & 10 \\ 
\hline 
 & $^3 T_{1g}$  & (9)  &   -25.702 & -4.238   &  -8.063   & -30.424   &  -11.151 \\ 
 & $^3 G_g$     & (12) &   -18.835 & -4.238   & -35.345   &  -2.015   &  -19.244 \\
 & $^3 T_{2g}$  & (9)  &  -10.987  & -39.553  &  11.212   & -28.746   &  -11.503 \\
 & $^1 G_g$     & (4)  &  -18.050  &   0.471 &  29.597   &  13.426   &   11.721 \\
 & $^1 A_g$     & (1)  &   74.162  &  59.330  &  57.274   &  74.228   &  136.987 \\
\hline
$*$ & $^1 H_g$   & (5)  & -18.492 &  0.471 & -12.595 &  29.970 &  6.970 \\
$*$ & $^1 H_g$   & (5)  &  55.180 & 16.951 &   8.429 & -18.442 &  64.255 
 
 \end{tabular} 
\end{ruledtabular}
\end{table} 
%
\begin{table} 
\caption{ 
Coefficients $\mu_l(p) \times 10^3$
for $(h_u^+)^3$. The coefficients
 marked by $(*)$ are not unique depending on integrals $v_l$, 
 see text for details.
\label{table6}     } 
\begin{ruledtabular}
 \begin{tabular}{ l c c r r r r r } 
  &  & deg. & $l=2$ & 4 & 6 & 8 & 10 \\ 
\hline
& $^4 G_u$    & (16) & -55.916 & -33.903 & -61.552 & -39.129 & -48.232 \\
& $^4 T_{1u}$ & (12) & -62.783 & -33.903 & -34.370 & -67.538 & -40.139 \\
& $^4 T_{2u}$ & (12) & -48.068 & -69.218 & -15.095 & -65.860 & -40.491 \\
& $^2 T_{1u}+{}^2T_{2u}$  & (12) &  -0.392 &  -5.650 & -30.218 & -13.676 &  30.095 \\
& $^2 T_{2u}$ &  (6) & -62.194 &  -5.650 &  18.710 & -11.293 &  -3.192 \\
& $^2 T_{1u}$ &  (6) & -47.480 & -40.966 &  37.985 &  -9.615 &  -3.544 \\
\hline
$*$ & $^2 H_u$  & (10) & -60.006 & -41.550 & -43.897 & -50.436 & -12.533 \\
$*$ & $^2 H_u$  & (10) & -51.236 &  -5.650 & -24.355 &  -5.511 & -10.208 \\
$*$ & $^2 H_u$  & (10) &  28.220 &   6.121 &  -1.637 &  24.146 &  61.563 \\
$*$ & $^2 H_u$  & (10) &  45.941 &  44.376 &  58.833 &  54.239 & 108.509 \\
$*$ & $^2 G_u$  &  (8) &  -2.237 & -40.966 & -36.070 & -43.844 &  38.009 \\
$*$ & $^2 G_u$  &  (8) &  38.730 &   6.121 &  10.229 & -40.794 &  60.323

  \end{tabular}
\end{ruledtabular}
\end{table}
Some coefficients depend on $v_l$. This did not occur
to $t_{1u}^2$ or $t_{1u}^3$, Table I.
These coefficients are marked by star ($*$)
in Tables~\ref{table5} and \ref{table6}. 
For them ($\mu_l^*$)
we give values which are calculated with only 
one parameter $v_l$. For example, $\mu_2^*$ corresponds to
the case when $v_{l=2} \neq 0$, while the others are zero, $v_{l \neq 2}=0$, 
and etc.
Interestingly, the stared terms have off-diagonal matrix elements
in the approach of Plakhutin {\it et al.}
(Tables 2 and 3 of Ref.~\onlinecite{Pla2}).
The appearance of $\mu_l^*$ implies that the computation
of energy splittings, Eq.~(\ref{2.1sub}), can not be separated in two
independent evaluations of $v_l$ and $\mu_l$. Finally, we remark
that the ``accidental" degeneracy of 
$^2 T_{1u}$ and ${}^2T_{2u}$ states of $(h_u)^3$
has been thoroughly studied in Refs.~\onlinecite{Oli,Lo,Jud,Pla2}.

\subsection{Excitonic configurations $h_u^+ t_{1u}^-$ and $h_u^+ t_{1g}^-$ }

In order to describe the excitonic configuration
$(h_u^+ t_{1u}^-)$ we introduce one $h_u$ hole 
(missing electron in HOMO) and one $t_{1u}$
electron. The basis functions read
\begin{eqnarray}
   | I \rangle =  | i_{hu}^+; \, i_u^- \rangle ,
\label{2.2sub} 
\end{eqnarray}
where now there are 10 states of the $h_u$ hole (index $i_{hu}^+$) 
and 6 states of the $t_{1u}$ electron (index $i_u^-$). 
The total number of basis functions
is 60. The important thing here is that {\it we have to treat
exchange differently}. If we consider $I$ as the initial state and  
$| J \rangle =  | j_{hu}^+; \, j_u^- \rangle$ as a final state, then
{\it the exchange transition $I \rightarrow J$ is described as
two electronic transitions $i_{u} \rightarrow i_{hu}$ and
$j_{u} \rightarrow j_{hu}$}. For the direct Coulomb interaction
we consider $i_{u} \rightarrow j_{u}$ and $j_{hu} \rightarrow i_{hu}$.
In addition, the sign of the direct Coulomb and exchange interactions
has to be reversed \cite{CS}. That is,
\begin{eqnarray}
   v_l(h_u^+ t_{1u}^-)=-v_l(h_u^- t_{1u}^-) ,
\label{2.3sub} 
\end{eqnarray}
for the direct Coulomb (even $l$)
and exchange matrices (even $l$ for $(h_u^+ t_{1u}^-)$ and odd $l$ for
$(h_u^+ t_{1g}^-)$).
The resulting energy spectrum for both configurations is given by
\begin{eqnarray}
  E_p   =  \triangle \epsilon + 
  \sum_{l} v_l^{ht}  \nu_l(p)  , 
\label{2.4sub} 
\end{eqnarray}
where $p=1-60$. Here the energy of the closed shell $(h_u)^{10}$ is
taken as zero and $\triangle \epsilon$ is an energy associated with
the promotion of one electron from the $h_u$ level 
to the $t_{1u}$ ($\triangle \epsilon=$2.7 eV \cite{Reed,Yang}) or 
$t_{1g}$ shell ($\triangle \epsilon=$3.85 eV \cite{Reed,Kato}) 
(see also Sec.\ V).
The calculated values of
$\nu_l(p)$ are quoted in Table~\ref{table4a} for $(h_u^+ t_{1u}^-)$
and in Table~\ref{table4} for $(h_u^+ t_{1g}^-)$. 
%
\begin{table} 
\caption{ 
 Coefficients $\nu_l(p) \times 10^3$ for $(h_u^+ t_{1u}^-)$.
\label{table4a}     } 
\begin{ruledtabular}
 \begin{tabular}{ c c r r r r r } 
     & deg. & $l=2$ & 4 & 6 & 8 & 10  \\ 
\hline 
$^3 H_g$    & (15) &  -6.474 &  21.189 & -0.381 &  11.652 & -23.332 \\
$^1 H_g$    & (5)  &  42.967 &  60.743 & 17.919 &  16.694 &  18.179 \\
$^3 G_g$    & (12) &  -8.240 &   0.0   & 11.184 & -10.435 &  19.526 \\
$^1 G_g$    & (4)  &  -8.240 &  14.126 & 13.725 &  24.304 & 103.205 \\
$^3 T_{1g}$ & (9)  &   6.474 & -21.189 &  9.701 & -11.652 &  38.081 \\
$^1 T_{1g}$ & (3)  &   6.474 & -21.189 & 93.578 & -11.652 &  70.528 \\
$^3 T_{2g}$ & (9)  &  15.303 & -14.126 & -0.678 &   6.145 &  11.644 \\
$^1 T_{2g}$ & (3)  &  15.303 & -14.126 & -0.678 &  70.660 &  69.707
 \end{tabular} 
\end{ruledtabular}
\end{table} 
%
\begin{table} 
\caption{ 
 Coefficients $\nu_l(p) \times 10^3$ for $(h_u^+ t_{1g}^-)$.
\label{table4}     } 
\begin{ruledtabular}
 \begin{tabular}{ c c r r r r r r} 
     & deg. & $l=2$ & 4 & 6 & 8 & 10 & \\ 
\hline 
$^3 G_u$    & (12)   &  21.425 &  0.0   & -19.130 & -5.219 &  18.167 &  \\
$^1 G_u$    & (4)    &  21.425 &  0.0   & -19.130 & -5.219 &  18.167 &  \\
$^3 H_u$    & (15)   &  16.834 &-16.204 &  20.434 &  4.128 &  12.320 &  \\
$^1 H_u$    &  (5)   &  16.834 &-16.204 &  20.434 &  4.128 &  12.320 &  \\
$^3 T_{1u}$ & (9)  & -16.834 & 16.204 & -14.058 & -4.128 & -27.802 &  \\
$^1 T_{1u}$ & (3)  & -16.834 & 16.204 & -14.058 & -4.128 & -27.802 &  \\
$^3 T_{2u}$ & (9)  & -39.789 & 10.802 &  21.449 &  4.207 & -55.659 &  \\
$^1 T_{2u}$ & (3)  & -39.789 & 10.802 &  21.449 &  4.207 & -55.659 &  \\
\hline
   & deg. & $l=1$ & 3 & 5 & 7 & 9 & 11 \\ 
\hline
$^3 G_u$    & (12) &  0.0   &  0.0  &  0.0   &  0.0   & 0.0   &  0.0   \\
$^1 G_u$    & (4)  &  0.0   &  0.0  &  0.0   & 19.382 &84.172 & 25.116 \\
$^3 H_u$    & (15) &  0.0   &  0.0  &  0.0   &  0.0   & 0.0   &  0.0   \\
$^1 H_u$    & (5)  &  0.0   &  0.0  &  0.0   & 45.811 & 3.826 & 34.493 \\
$^3 T_{1u}$ & (9)  &  0.0   &  0.0  &  0.0   &  0.0   & 0.0   &  0.0   \\
$^1 T_{1u}$ & (3)  & 91.820 &  0.0  & 83.877 & 13.282 & 0.261 & 63.149 \\
$^3 T_{2u}$ & (9)  &  0.0   &  0.0  &  0.0   &  0.0   & 0.0   &  0.0   \\
$^1 T_{2u}$ & (3)  &  0.0   &142.831& 21.746 &  5.520 & 0.662 &106.374
 \end{tabular} 
\end{ruledtabular}
\end{table} 
The coefficients $\nu_l(p)$ in the Tables correspond
to $v_l^{ht}$ with the plus sign, i.e. $v_{l}^{ht}=v_l > 0$,
where $v_l$ are given by Eq.~(\ref{2.10}).

\subsection{($t_{1u} t_{1g}$) configuration}

The basis functions here are
\begin{eqnarray}
   | I \rangle =  | i_u; \, i_g \rangle ,
\label{2.10sub} 
\end{eqnarray}
where, as before, the index $i_u$ stands for the $t_{1u}$ 
LUMO-level ($i_u=1-6$), while
the index $i_g=(k,s_z)$ stands for the three $t_{1g}$ 
(LUMO+1)-orbitals, Eqs.~(\ref{1.3a}-c),
and the spin projection $s_z$. Thus, $i_g=1-6$ and
in total, there are
$6 \times 6=36$ nonequivalent basis vectors $| I \rangle$, Eq.\ (\ref{2.1}). 
From the theory
of addition of angular momenta and selection rules
we deduce that for the direct Coulomb
interactions only the even values $l=0$, 2, 4, 6, 8 and 10 are relevant,
while for exchange these are odd numbers from one to eleven.
The calculated $\lambda_l(p)$ are quoted in Table~\ref{table3}.
%
\begin{table} 
\caption{ 
 Coefficients $\lambda_l(p) \times 10^3$ for 
 $(t_{1u} t_{1g})$; $\lambda_3=0$.
\label{table3}     } 
\begin{ruledtabular}
 \begin{tabular}{ c c r r r r r } 
     & deg. & $l=2$ & 4 & 6 & 8 & 10 \\ 
\hline 
$^1 A_u$ & (1)  & -15.303 & -32.407 &  5.797 & -22.605 & -15.059 \\
$^3 A_u$ & (3)  & -15.303 & -32.407 &  5.797 & -22.605 & -15.059 \\
$^3 H_u$ & (15) &  -1.530 &  -3.241 & 14.927 &  -2.260 &  -4.757 \\
$^1 H_u$ & (5)  &  -1.530 &  -3.241 & 14.927 &  -2.260 &  -4.757 \\
$^3 T_{1u}$ & (9)  &   7.652 &  16.204 & 21.014 &  11.302 &   2.111 \\
$^1 T_{1u}$ & (3)  &   7.652 &  16.204 & 21.014 &  11.302 &   2.111 \\
\hline
  & deg. & $l=1$ & 5 & 7 & 9 & 11 \\ 
\hline
$^1 A_u$ & (1)  & -64.274 &  41.939 & -13.747 &  31.956 &   4.127 \\
$^3 A_u$ & (3)  &  64.274 & -41.939 &  13.747 & -31.956 &  -4.127 \\
$^3 H_u$ & (15) & -32.137 &  -4.194 &  -8.442 & -14.152 & -23.791 \\
$^1 H_u$ & (5)  &  32.137 &   4.194 &   8.442 &  14.152 &  23.791 \\
$^3 T_{1u}$ & (9)  & -32.137 & -20.969 &  -9.487 & -34.238 & -41.028 \\
$^1 T_{1u}$ & (3)  &  32.137 &  20.969 &   9.487 &  34.238 &  41.028 
 \end{tabular} 
\end{ruledtabular}
\end{table} 

\subsection{Three electron configuration $(t_{1u})^2 t_{1g}$}

In case of the three-electron $(t_{1u})^2 t_{1g}$ configuration,
we construct $(6\times5/2)\times6=90$ basis vectors 
\begin{eqnarray}
 | I' \rangle =| i_{u1} i_{u2} i_g \rangle . 
\label{2.5.new} 
\end{eqnarray}
Here $i_{u1}$ and $i_{u2}$ are indices of the $t_{1u}$ LUMO states,
{\it i.e.} $i_{u1},i_{u2}=1-6$ and $i_g$ labels six $t_{1g}$ states
($i_g=1-6$). Since in that case we deal with two equivalent
$t_{1u}$ electrons, the basis functions are taken with $i_{u1} > i_{u2}$. 
The calculated $\lambda_l$ ($p=1-90$) are quoted in Table~\ref{table9}.
%
\begin{table} 
\caption{ 
 Coefficients $\lambda_l(p) \times 10^3$ for 
 $(t_{1u})^2 t_{1g}$; $\lambda_3=0$. The coefficients
 marked by $(*)$ are not unique 
 depending on integrals $v_l$, see text for details.
\label{table9}     } 
\begin{ruledtabular}
 \begin{tabular}{l c c r r r r r} 
 &  & deg. & $l=2$ & 4 & 6 & 8 & 10 \\ 
\hline 
& $^4 H_g$    & (20)    &  -1.413 & -17.949 & 54.714 & -37.446 & -43.636 \\
& $^2G_g+{}^2T_{2g}$ & (14)  &  -2.472 &  -2.244 & 53.450 &   3.420 &   2.127 \\
& $^4 T_{1g}$ & (12)    & -10.595 & -37.393 & 48.627 & -51.009 & -50.504 \\
& $^4 A_g$    &  (4)    &  12.360 &  11.218 & 63.844 & -17.102 & -33.333 \\
& $^2 A_g$    &  (2)    &  12.360 &  11.218 & 63.844 & -17.102 & -33.333 \\
\hline
$*$ & $^2 H_g$   &(10) &  -1.413 & -17.949 & 54.714 & -37.446 & -43.636 \\
$*$ & $^2 H_g$   &(10) &  11.301 &  26.923 & 62.580 &  23.765 &  12.429 \\
$*$ & $^2 T_{1g}$ & (6) & -17.976 & -30.048 & 42.958 & -12.337 &  -4.859 \\
$*$ & $^2 T_{1g}$ & (6) &  13.738 &  53.979 & 63.568 &  83.868 &  79.472 \\
$*$ & $^2 T_{1g}$ & (6) &  -1.413 & -17.949 & 54.714 & -37.446 & -43.636 \\
\hline
\hline
 &  & deg. & $l=1$ & 5 & 7 & 9 & 11 \\ 
\hline
& $^4 H_g$    & (20)    & -64.274 & -16.775 & -17.407 & -38.347 & -56.201 \\
& $^2G_g+{}^2T_{2g}$ & (14)    & -32.137 &  -4.194 &  -8.442 & -14.152 & -23.791 \\
& $^4 T_{1g}$ & (12)    &   0.0   & -41.939 &  -2.614 & -50.216 & -43.092 \\
& $^4 A_g$    &  (4)    & -64.274 & -41.939 & -18.975 & -68.476 & -82.056 \\
& $^2 A_g$    &  (2)    &  32.137 &  20.969 &   9.487 &  34.238 &  41.028 \\
\hline
$*$ & $^2 H_g$    &(10)  & -32.137 & -21.987 &  -9.260 & -33.523 & -39.202 \\
$*$ & $^2 H_g$    &(10)  &  32.137 &  13.599 &   8.738 &  23.480 &  30.584 \\ 
$*$ & $^2 T_{1g}$ & (6)  & -64.274 & -41.939 & -14.929 & -33.480 & -40.641 \\
$*$ & $^2 T_{1g}$ & (6)  & -32.137 & -17.803 &  -9.006 & -26.724 & -18.081 \\
$*$ & $^2 T_{1g}$ & (6)  &  96.411 &  34.578 &  22.366 &  30.075 &  32.867 
 \end{tabular} 
\end{ruledtabular}
\end{table} 
The important peculiarity of $\lambda_l$ is
that as in the case of few holes, Sec.\ IV-A,
the coefficients marked by star ($*$) in Table~\ref{table9}
(for two $^2H_g$ and three $^2 T_u$ terms), exhibit dependence on $v_l$. 
For these coefficients ($\lambda_l^*$)
we give values which are calculated with only 
one parameter $v_l$. For example, $\lambda_2^*$ corresponds to
a calculation where $v_{l=2} \neq 0$, $v_{l \neq 2}=0$ and etc.

Another very interesting observation is that the energy of 
the $^2G_g$ term ``accidentally" coincides with the $^2T_{g}$ states, 
Table~\ref{table9}.
The same feature has been found for the $t_{1u}(t_{1g})^2$ configuration.

\section {Energy levels} 
\label{sec:en} 

In order to study the splitting quantitatively we calculated
the integrals $v_l$ using three models for radial dependence ${\cal R}$
of $t_{1u}$ MOs. In the first model (I) we assume a delta
dependence, $i.e.$
\begin{eqnarray}
 {\cal R}(r)=\delta(r-r_{C_{60}})/r^2 ,
\label{2.16} 
\end{eqnarray}
where $r_{C_{60}}=3.55$ {\AA} is the radius of the C$_{60}$ molecule.
This gives
\begin{eqnarray}
 v_l=\frac{4\pi}{2l+1} \frac{1}{r_{C_{60}}} .
\label{2.16a} 
\end{eqnarray}

In the second model (II), Fig.\ 2, we use 
\begin{eqnarray}
 {\cal R}(r)=C \exp(-\sqrt{2 |E|}\,|r-r_{C_{60}}|) ,
\label{2.17} 
\end{eqnarray}
where $C$ is determined from the normalization condition
and $E=-5.863$ eV, the energy of the carbon $p_z$ atomic orbital 
in atomic calculations in local density approximation (LDA)
(such dependence corresponds to the large distance
limit for the carbon $p_z$-orbital).
%
\begin{figure} 
\resizebox{0.46\textwidth}{!}
{ 
 \includegraphics{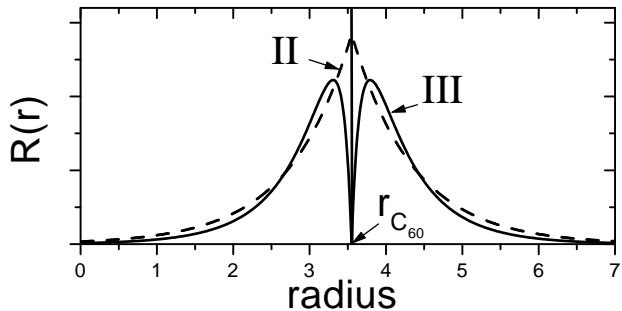} 
} 
\vspace{2mm}
\caption{
$|{\cal R}(r)|$ for model II (dashed line) and model III (full line)
 as a function of $r$ (in {\AA}).} 
\label{fig2} 
\end{figure} 
%
In the third model (III) we take
\begin{eqnarray}
 {\cal R}(r)=C' {\cal R}_{p_z}(|r-r_{C_{60}}|) ,
\label{2.18} 
\end{eqnarray}
where $C'$ is a normalization constant and ${\cal R}_{p_z}$
is the radial dependence of the carbon $p_z$ orbital in LDA, Fig.~2.
The calculated values of $v_l$ are quoted in Table~\ref{table10}.
%
\begin{table} 
\caption{ 
 Calculated $v_l$ for models I, II and III; in eV.
\label{table10}     } 
\begin{ruledtabular}
 \begin{tabular}{c c c r r r r } 
 model & $l=2$ & 4 & 6 & 8 & 10 & 12 \\ 
 I  & 10.195 & 5.664 & 3.921 & 2.998 & 2.427 & 2.039 \\ 
 II &  6.919 & 3.064 & 1.752 & 1.137 & 0.797 & 0.589 \\
 III&  6.798 & 2.965 & 1.673 & 1.074 & 0.747 & 0.550 \\
\hline 
    & $l=1$  & 3 & 5 & 7 & 9 & 11 \\
 I  & 16.991 & 7.282 & 4.634 & 3.398 & 2.683 & 2.216 \\
 II & 13.145 & 4.390 & 2.270 & 1.395 & 0.944 & 0.681 \\
 III& 13.018 & 4.279 & 2.181 & 1.324 & 0.889 & 0.637 
 \end{tabular} 
\end{ruledtabular}
\end{table} 

First, from Eqs.\ (\ref{2.15}) and (\ref{2.20}) we calculate 
the molecular terms of $t_{1u}^2$ and $t_{1u}^3$, Table \ref{table2}.
%
\begin{table} 
\caption{ 
 Molecular terms and their degeneracies 
 (in parentheses) for $(t_{1u})^2$ and $(t_{1u})^3$
  calculated with models I, II and III; in eV.
  $({}^n_2)U_0$ is zero of energy.
\label{table2}     } 
\begin{ruledtabular}
 \begin{tabular}{c | r r r | r r r } 
    & \multicolumn{3}{c|}{$(t_{1u})^2$} &
      \multicolumn{3}{c}{$(t_{1u})^3$} \\ 
    & $^3T_{1g}$  (9) & $^1H_g$  (5) & $^1A_g$  (1) 
    &  $^4 A_u$  (4) & $^2 H_u$  (10) & $^2T_{1u}$  (6) \\
\hline
 I  &  -0.275 & 0.175 & 0.848 &  -0.824 & -0.150 & 0.299 \\ 
 II &  -0.122 & 0.077 & 0.375 &  -0.366 & -0.068 & 0.131 \\
 III&  -0.117 & 0.073 & 0.359 &  -0.351 & -0.066 & 0.125
 \end{tabular} 
\end{ruledtabular}
\end{table} 
We observe that despite their differences, models II and III give very close
values. Therefore, we believe that the calculated parameters $v_l$ of models
II and III are realistic and 
will change little if a more refined calculation of ${\cal R}$ is made.
On the other hand, the first model is a rude approximation, 
and in the following we will not use it. 
Notice that the $t_{1u}^2$ and $t_{1u}^3$ energy spectra are the
analogue of $p^2$ ($^3 P$, $^1 D$, $^1 S$) and $p^3$ ($^4 S$, $^2 D$, $^2 P$)
terms in atomic physics \cite{CS}.
This occurs because $p_x$, $p_y$ and $p_z$ orbitals
($Y_1^{1,c}$, $Y_1^{1,s}$ and $Y_1^0$) also belong
to the $t_{1u}$ irrep of $I_h$ \cite{Coh}.
In Table \ref{table2} as well as in all other Tables of this section
(\ref{table_14}-\ref{table_12}) the energy associated with the
spherically symmetric multipole component (i.e. $({}^n_2)U_0$ for
electrons, $({}^{10-m}_{\;\;\;2})U_0$ for holes and $\triangle \epsilon$ for
excitonic configurations) is put to zero.

Similarly, one can obtain the energy levels of
$t_{1g}^2$ and $t_{1g}^3$, Table \ref{table_14}.
%
\begin{table} 
\caption{ 
 Molecular terms of $(t_{1g})^2$ and $(t_{1g})^3$
 calculated with models II and III; in eV.
 $({}^n_2)U_0$ is zero of energy.
\label{table_14}     } 
\begin{ruledtabular}
 \begin{tabular}{c | r r r | r r r } 
    & \multicolumn{3}{c|}{$(t_{1g})^2$} &
      \multicolumn{3}{c}{$(t_{1g})^3$} \\ 
    & $^3 T_{1g}$ & $^1 H_g$ & $^1 A_g$ & $^4 A_g$ & $^2 H_g$ & $^2 T_{1g}$ \\ 
    & (9) & (5) & (1) &  (4) & (10) & (6) \\
\hline
 II &  -0.221 & 0.086 & 0.547 &  -0.664 & -0.203 & 0.104 \\
 III&  -0.215 & 0.083 & 0.529 &  -0.646 & -0.199 & 0.099
 \end{tabular} 
\end{ruledtabular}
\end{table} 
Notice that the energy span of $t_{1u}^2$ and $t_{1u}^3$
configurations is almost the same, $\sim0.5$ eV,
and it is smaller than that of $t_{1g}^2$ and $t_{1g}^3$ states,
$\sim0.75$ eV.
However, the zero of energy in Tables \ref{table2} and \ref{table_14}
 is different for $t_{1u}^n$ and $t_{1g}^n$ configurations.
When one $t_{1u}$ electron is promoted to a $t_{1g}$ state, its energy 
is increased by $\sim$1.153 eV \cite{Reed,Kato}, that is,
\begin{eqnarray}
  \triangle \epsilon_1 =\epsilon(t_{1g})-\epsilon(t_{1u}) 
  \approx 1.153 \mbox{ eV} .
\label{2.18a} 
\end{eqnarray}
This one-electron energy difference accounts for the interaction of
the electron with the carbon nuclei and the ``core" like $\sigma-$ and
$\pi-$ electrons. Therefore, comparing the energy of $(t_{1g})^2$ with
that of $(t_{1u})^2$ states one should add $2 \triangle \epsilon_1$ to
the $(t_{1g})^2$ values. For the case of three electrons ($(t_{1u})^3$ and
$(t_{1g})^3$) we add $3 \triangle \epsilon_1$ to the $(t_{1g})^3$ values.

%
\begin{table} 
\caption{ 
 Molecular terms of $(h_u^+)^2$ 
 and $(h_u^+)^3$ calculated with models II and III; in eV.
 $({}^{10-m}_{\;\;\;2})U_0$ is zero of energy.
\label{table_13}     } 
\begin{ruledtabular}
 \begin{tabular}{c c r r | c c r r} 
     \multicolumn{4}{c|}{$(h_u^+)^2$} &
      \multicolumn{4}{c}{$(h_u^+)^3$} \\
\hline 
  & deg. & II & III &   & deg. &  II & III \\
\hline
$^3 T_{1g}$& (9)  & -0.248 & -0.242 & $^4 T_{1u}$& (12) & -0.707 & -0.687 \\
$^3 G_g$   & (12) & -0.223 & -0.216 & $^4 G_u$   & (16) & -0.682 & -0.662 \\
$^3 T_{2g}$& (9)  & -0.219 & -0.213 & $^4 T_{2u}$& (12) & -0.678 & -0.658 \\
$^1 G_g$   & (4)  & -0.047 & -0.049 & $^2 T_{2u}$& (6)  & -0.430 & -0.423 \\
$^1 H_g$   & (5)  & -0.040 & -0.042 & $^2 H_u$   & (10) & -0.414 & -0.407 \\
$^1 H_g$   & (5)  &  0.410 &  0.402 & $^2 T_{1u}$& (6)  & -0.401 & -0.394 \\
$^1 A_g$   & (1)  &  0.989 &  0.958 & $^2 H_u$   & (10) & -0.396 & -0.389 \\
           &      &        &        & $^2 G_u$   & (8)  & -0.077 & -0.075 \\
           &      &        &        & $^2 T_{1u}+{}^2 T_{2u}$ & (12) & -0.065 & -0.062 \\
           &      &        &        & $^2 H_u$   & (10) &  0.156 &  0.155 \\
           &      &        &        & $^2 G_u$   & (8)  &  0.160 &  0.159 \\
           &      &        &        & $^2 H_u$   & (10) &  0.531 &  0.514

 \end{tabular} 
\end{ruledtabular}
\end{table} 
%
\begin{table} 
\caption{ 
 Molecular terms of $(h_u^+)^4$ 
 and $(h_u^+)^5$ calculated with models II and III; in eV.
 $({}^{10-m}_{\;\;\;2})U_0$ is zero of energy.
\label{table_14+}     } 
\begin{ruledtabular}
 \begin{tabular}{c c r r | c c r r} 
     \multicolumn{4}{c|}{$(h_u^+)^4$} &
      \multicolumn{4}{c}{$(h_u^+)^5$} \\
\hline 
  & deg. & II & III &   & deg. &  II & III \\
\hline
$^5 H_g$    & (25) & -1.377 & -1.337 & $^6 A_u$    & (6)  & -2.294 & -2.228 \\
$^3 G_g$    & (12) & -1.005 & -0.982 & $^4 H_u$    & (20) & -1.716 & -1.672 \\
$^3 H_g$    & (15) & -1.004 & -0.981 & $^2 H_u$    & (10) & -1.324 & -1.294 \\
$^1 G_g$    & (4)  & -0.819 & -0.805 & $^2 G_u$    & (8)  & -1.323 & -1.293 \\
$^1 A_g$    & (1)  & -0.814 & -0.799 & $^4 H_u$    & (20) & -1.265 & -1.228 \\
$^3 T_{1g}$ & (9)  & -0.787 & -0.767 & $^4 G_u$    & (16) & -1.258 & -1.221 \\
$^3 G_g$    & (12) & -0.780 & -0.760 & $^2 A_u$    & (2)  & -1.106 & -1.079 \\
$^3 T_{2g}$ & (9)  & -0.766 & -0.747 & $^2 G_u$    & (8)  & -1.094 & -1.067 \\
$^1 H_g$    & (5)  & -0.491 & -0.482 & $^4 T_{2u}$ & (12) & -1.086 & -1.057 \\
$^1 T_{2g}$ & (3)  & -0.490 & -0.481 & $^4 G_u$    & (16) & -1.083 & -1.054 \\
$^1 H_g$    & (5)  & -0.570 & -0.460 & $^4 T_{1u}$ & (12) & -1.057 & -1.028 \\
$^1 T_{1g}$ & (3)  & -0.461 & -0.451 & $^2 H_u$    & (10) & -0.816 & -0.800 \\
$^3 T_{2g}$ & (9)  & -0.445 & -0.430 & $^2 T_{1u}$ & (6)  & -0.803 & -0.787 \\
$^3 H_g$    & (15) & -0.440 & -0.426 & $^2 H_u$    & (10) & -0.792 & -0.775 \\
$^3 T_{1g}$ & (9)  & -0.434 & -0.419 & $^2 T_{2u}$ & (6)  & -0.771 & -0.755 \\
$^3 H_g$    & (15) & -0.434 & -0.419 & $^2 H_u$    & (10) & -0.769 & -0.748 \\
$^3 G_g$    & (12) & -0.160 & -0.157 & $^2 T_{2u}$ & (6)  & -0.769 & -0.747 \\
$^3 T_{2g}$ & (9)  & -0.160 & -0.157 & $^2 H_u$    & (10) & -0.747 & -0.726 \\
$^3 T_{1g}$ & (9)  & -0.150 & -0.148 & $^2 T_{1u}$ & (6)  & -0.740 & -0.718 \\
$^1 A_g$    & (1)  & -0.148 & -0.143 & $^2 T_{1u}$ & (6)  & -0.487 & -0.476 \\
$^1 G_g$    & (4)  & -0.138 & -0.134 & $^2 G_u$    & (8)  & -0.453 & -0.442 \\
$^1 H_g$    & (5)  &  0.020 &  0.014 & $^2 T_{2u}$ & (6)  & -0.432 & -0.421 \\
$^1 G_g$    & (4)  &  0.041 &  0.035 & $^2 G_u$    & (8)  & -0.417 & -0.401 \\
$^1 H_g$    & (5)  &  0.087 &  0.089 & $^2 A_u$    & (2)  & -0.411 & -0.395 \\
$^1 G_g$    & (4)  &  0.090 &  0.091 & $^2 G_u$    & (8)  & -0.231 & -0.223 \\
$^1 H_g$    & (5)  &  0.484 &  0.471 & $^2 H_u$    & (10) & -0.227 & -0.218 \\
$^1 A_g$    & (1)  &  1.062 &  1.027 & $^2 H_u$    & (10) &  0.144 &  0.135

 \end{tabular} 
\end{ruledtabular}
\end{table} 
Next, in Tables \ref{table_13} and \ref{table_14+} we give results
for the hole configurations $(h_u^+)^m$.
We observe that the energy span of $(h_u^+)^2$ and $(h_u^+)^3$, $\sim1.2$ eV,
is almost the same. The magnitude is larger than for electronic $t_{1u}^n$ and
$t_{1g}^n$ configurations, Tables \ref{table2} and \ref{table_14}. 
Even a larger value of energy splitting, 
$\sim 2.4$ eV,
was obtained for the case of four and five holes, $(h_u^+)^4$ and $(h_u^+)^5$.
Another important observation is that the number of states in 
a small energy interval
$\triangle \varepsilon \sim 0.03$ eV near the ground state
is 30, 40, 25, 6 for $m=$2, 3, 4, 5, respectively.
This suggests that the configuration of $m=3$ holes is
most susceptible for Jahn-Teller distortions of the C$_{60}$
molecule and hence for hole-phonon coupling which
causes superconductivity \cite{Bat1}.

The results of calculations of excitonic configurations 
$(h_u^+ t_{1u}^-)$ and $(h_u^+ t_{1g}^-)$ are quoted in Table \ref{table_12a}.  
%
\begin{table} 
\caption{ 
 Molecular terms of the excitonic configurations $(h_u^+ t_{1u}^-)$
 and $(h_u^+ t_{1g}^-)$ calculated with models II and III; in eV.
  $\triangle \epsilon$ in Eq.~(\ref{2.4sub}) is zero of energy. 
\label{table_12a}     } 
\begin{ruledtabular}
 \begin{tabular}{c c r r | c c r r} 
     \multicolumn{4}{c|}{$(h_u^+ t_{1u}^-)$} &
      \multicolumn{4}{c}{$(h_u^+ t_{1g}^-)$} \\
\hline 
    & deg. & II & III &   & deg. &  II & III \\
\hline
$^3 G_g$    & (12) & -0.034 & -0.034 & $^3 T_{2u}$ & (9)  & -0.244 & -0.240 \\
$^3 T_{1g}$ & (9)  &  0.014 &  0.013 & $^3 T_{1u}$ & (9)  & -0.118 & -0.115 \\
$^3 H_g$    & (15) &  0.014 &  0.013 & $^3 H_u$    & (15) &  0.117 &  0.114 \\
$^3 T_{2g}$ & (9)  &  0.078 &  0.076 & $^3 G_u$    & (12) &  0.123 &  0.122 \\
$^1 G_g$    & (4)  &  0.120 &  0.112 & $^1 H_u$    & (5)  &  0.208 &  0.200 \\
$^1 T_{1g}$ & (3)  &  0.187 &  0.178 & $^1 G_u$    & (4)  &  0.247 &  0.238 \\
$^1 T_{2g}$ & (3)  &  0.197 &  0.189 & $^1 T_{2u}$ & (3)  &  0.513 &  0.495 \\
$^1 H_g$    & (5)  &  0.548 &  0.534 & $^1 T_{1u}$ & (3)  &  1.341 &  1.321

 \end{tabular} 
\end{ruledtabular}
\end{table} 
The energy span of $(h_u^+ t_{1g}^-)$, $\sim 1.5$ eV, greatly exceeds that
of $(h_u^+ t_{1u}^-)$, $\sim 0.57$ eV.
A promotion of one electron to the $t_{1u}$ shell
increases the one-electron energy by the factor 
\begin{eqnarray}
  \triangle \epsilon_2 =\epsilon(t_{1u})-\epsilon(h_u) 
  \approx 2.69 \mbox{ eV} .
\label{2.18b} 
\end{eqnarray}
The quantity $\triangle \epsilon_2$ 
is called electron affinity of C$_{60}$ and it was measured
experimentally \cite{Yang,Reed}. 
It accounts for the energy difference due to the interactions of the electron
with the carbon nuclei and the ``core" electrons. The value should
be taken into account in Eq.~(\ref{2.4sub}) 
(i.e. $\triangle \epsilon=\triangle \epsilon_2$) when $(h_u^+ t_{1u}^-)$
is compared with the ground state energy of the neutral molecule.
In the case of $(h_u^+ t_{1g}^-)$ one should use
$\triangle \epsilon=\triangle \epsilon_1+\triangle \epsilon_2=3.85$ eV
in Eq.~(\ref{2.4sub}). 

The molecular terms for $(t_{1u} t_{1g})$ and $(t_{1u})^2 t_{1g}$ 
are given in Table \ref{table_12}.
%
\begin{table} 
\caption{ 
 Molecular terms of $(t_{1u} t_{1g})$ 
 and $(t_{1u})^2 t_{1g}$ calculated with models II and III; in eV.
 $^2 G_g$ and $^2 T_{2g}$ are ``accidentally" degenerate.
 $({}^n_2)U_0$ is zero of energy.
\label{table_12}     } 
\begin{ruledtabular}
 \begin{tabular}{c c r r | c c r r} 
     \multicolumn{4}{c|}{$(t_{1u} t_{1g})$} &
      \multicolumn{4}{c}{$(t_{1u})^2 t_{1g}$} \\
\hline 
    & deg. & II & III &   & deg. &  II & III \\
\hline
$^1 A_u$   & (1)  & -0.969 & -0.958 & $^4 H_g$    & (20) & -1.028 & -1.010 \\
$^3 H_u$   & (15) & -0.474 & -0.467 & $^2 T_{1g}$ & (6)  & -0.991 & -0.979 \\
$^3 T_{1u}$& (9)  & -0.390 & -0.384 & $^4 A_g$    & (4)  & -0.901 & -0.886 \\
$^1 H_u$   & (5)  &  0.473 &  0.465 & $^2 G_g+{}^2 T_{2g}$ & (14) & -0.398 & -0.395 \\
$^3 A_u$   & (3)  &  0.503 &  0.506 & $^4 T_{1g}$ & (12) & -0.376 & -0.361 \\
$^1 T_{1u}$& (3)  &  0.697 &  0.682 & $^2 H_g$    & (10) & -0.223 & -0.224 \\
           &      &        &        & $^2 T_{1g}$ & (6)  &  0.004 & -0.008 \\
           &      &        &        & $^2 H_g$    & (10) &  0.449 &  0.443 \\
           &      &        &        & $^2 A_g$    & (2)  &  0.729 &  0.714 \\
           &      &        &        & $^2 T_{1g}$ & (6)  &  1.052 &  1.045

 \end{tabular} 
\end{ruledtabular}
\end{table} 
The energy span of the excited configurations is relatively large.
It is approximately 1.6 eV for $(t_{1u} t_{1g})$ and 2 eV for
$(t_{1u})^2 t_{1g}$. 
Since both configurations imply the excitation of one $t_{1u}$ electron
to a $t_{1g}$ state,
 we should add $\triangle \epsilon_1$
to the energies of the  $(t_{1u} t_{1g})$ and $(t_{1u})^2 t_{1g}$
molecular terms, 
when we compare them with those of the $(t_{1u})^2$ and $(t_{1u})^3$
configurations.
The $(t_{1u})^2$ and $(t_{1u} t_{1g})$ 
(as well as $(t_{1u})^3$ and $(t_{1u})^2 t_{1g}$)
groups of terms are of different parity and thus
there is no configuration mixing between them.
Therefore, although some two-electron molecular terms
of the $(t_{1u})^2$ and $(t_{1u} t_{1g})$ configurations overlap,
they do not interact with each other.
The same holds for the $(t_{1u})^3$ and $(t_{1u})^2 t_{1g}$ configurations.

However, there can be a hybridization between terms
of the same symmetry of $(t_{1u})^2$ and 
$(t_{1g})^2$ configurations. The $(t_{1g})^2$ configuration requires
promotion of two electrons to the $t_{1g}$ shell,
with a subsequent energy increase of
$2\triangle \epsilon_1 \sim 2.3$ eV. 
Since the value is relatively large,
the hybridization is expected to be weak.
In order to study this issue we have carried out calculations 
where the mixing between the $(t_{1u})^2$ and $(t_{1g})^2$ configurations
was allowed. In the calculation we have considered couplings
between two $^3 T_{1g}$ levels, two $^1 H_g$ levels and two
$^1 A_g$ levels at different values of $\triangle \epsilon_1$. 
As before, we have employed the method described in Sec.~III.
We have found that
the energy spectrum separates in two groups. A
group at lower energies originates from the former $(t_{1u})^2$
levels, while the other group at higher energies has a large
parentage of the $(t_{1g})^2$ states.
In Fig.~3 we plot the energies of the three lowest levels as a function
of $\triangle \epsilon_1$.
%
\begin{figure} 
\resizebox{0.46\textwidth}{!}
{ 
 \includegraphics{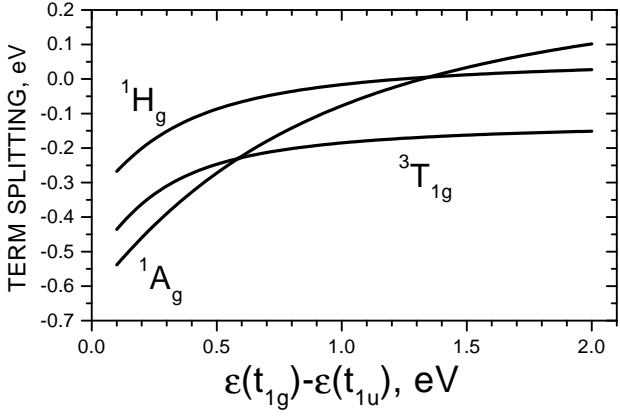} 
} 
\vspace{2mm}
\caption{
Three lowest levels of the coupled $(t_{1u})^2+(t_{1g})^2$
configurations as a function of 
$\triangle \epsilon_1 =\epsilon(t_{1g})-\epsilon(t_{1u})$.
The ground state is the $^1 A_g$ singlet for $\triangle \epsilon_1 < 0.58$~eV,
and the $^3 T_{1g}$ triplet for $\triangle \epsilon_1 > 0.58$~eV.
} 
\label{fig3} 
\end{figure} 
%
An interesting feature of Fig.~3 is the crossing of the $^3 T_{1g}$ triplet 
with the $^1 A_g$ singlet at 0.58~eV with subsequent
inversion of their positions.
Thus, if $\triangle \epsilon_1 < 0.58$~eV then the ground state is
the $^1 A_g$ singlet, while for $\triangle \epsilon_1 > 0.58$~eV
the ground state is the $^3 T_{1g}$ triplet.
This unusual behavior explains why the $^1 A_g$ singlet was
reported as the ground state of C$_{60}^{2-}$ 
by Negri {\it et al.}, Ref.~\onlinecite{Neg} (QCFF/$\pi$ method).
From our calculation (Fig.~3) it follows that the reason for this is
a small energy difference 
between $t_{1g}$ and $t_{1u}$ states.
In Ref.~\onlinecite{Neg} $\triangle \epsilon_1$=0.64 eV, 
which is only half 
of the experimental value of 1.153 eV for C$_{60}^-$ \cite{Reed}. 
Such low lying $t_{1g}$ states lead to
an overestimation of the $^1 A_g\,[t_{1u}^2]\,-\,^1 A_g\,[t_{1g}^2]$ 
configuration mixing and lowering of the bonding $^1 A_g$ term 
below $^3 T_{1g}$. 
The experimental value $\triangle \epsilon_1$=1.153~eV \cite{Kato,Reed}
implies that the ground state is a triplet, as obtained by
our calculations and in accordance with Hund's rules.

\section {Magnetic moments} 
\label{sec:mm} 

In this section we will calculate the magnetic moments
of C$_{60}^{n\pm}$ 
for different orientations of the molecule.
In a small external magnetic field $\vec{H}$
we add to a many body Coulomb interaction
$V(\vec{r},\vec{r}\,')$, Eq.~(\ref{2.7}), a magnetic term
\begin{eqnarray}
   V_{mag} = - {\cal M}_z \cdot H ,
\label{m.1} 
\end{eqnarray}
where ${\cal M}_z=\sum_{k=1}^n M_z(k)$ is  a sum of one-electron (one-hole)
terms with
\begin{eqnarray}
   \vec{M}(k) = \mu_B (\vec{L}(k) + 2 \vec{S}(k)) . 
\label{m.2} 
\end{eqnarray}
Here $\mu_B$ is the Bohr magneton, $k=1-n$ for electrons and
$k=1-m$ for holes. 
The magnetic moment (\ref{m.1}) belongs to the class of
one-particle operators discussed in Sec.~III.
Explicitly, for the two-particle case
we find
\begin{eqnarray}
& &  \langle I | V_{mag} | J \rangle =
 [\langle i_1 | M_z | j_1 \rangle\, \delta(i_2 j_2) 
  +  \langle i_2 | M_z | j_2 \rangle\, \delta(i_1 j_1) \nonumber \\
  & & -\langle i_1 | M_z | j_2 \rangle\, \delta(i_2 j_1) 
  -  \langle i_2 | M_z | j_1 \rangle\, \delta(i_1 j_2)
  ] \cdot H ,  
\label{m.3} 
\end{eqnarray}
where $\langle i | M_z | j \rangle$ 
stands for the one-particle matrix elements.
The generalization of the procedure 
for a many particle case is given in Sec.~III.

The one-particle matrix elements of
spin momentum are given by the standard expressions \cite{Bra,Tin}.
They are independent of the orientation of the C$_{60}$ molecule. 
In order to calculate the orbital polarization we start
with the C$_{60}$ molecule in the orientation of Cohan \cite{Coh}.
By means of Eqs.\ (\ref{1.2a}-c) and taking into account that
\begin{eqnarray}
  L_z\, Y_l^{m,c}=i\,m Y_l^{m,s},\;\;\;\; 
  L_z\, Y_l^{m,s}=-i\,m Y_l^{m,c} ,
 \label{5.6} 
\end{eqnarray}
we obtain for the orbital momentum of $t_{1u}$ states 
\begin{subequations}
\begin{eqnarray}
 & & \langle t_{1u},2 | L_z | t_{1u},3 \rangle=2.5 i,  \label{5.7a}  \\
 & & \langle t_{1u},3 | L_z | t_{1u},2 \rangle=-2.5 i .  \label{5.7b} 
\end{eqnarray}
\end{subequations}
The other matrix elements of $L_z$ are zero.
(Here and below all values of magnetic moments are given in $\mu_B$.)
In the case of $h_u$ states we find
\begin{subequations}
\begin{eqnarray}
  \langle h_u,2 | L_z | h_u,3 \rangle=\frac{1}{2} i, \quad
   \langle h_u,3 | L_z | h_u,2 \rangle=-\frac{1}{2} i,  
 \quad \quad \label{5.8a}  \\
  \langle h_u,4 | L_z | h_u,5 \rangle= i , \quad
  \langle h_u,5 | L_z | h_u,4 \rangle=-i , \quad \quad \quad \label{5.8b} 
\end{eqnarray}
\end{subequations}
and the rest is zero.

There are two equivalent approaches to study the C$_{60}$
molecule in the magnetic field. The first is to use an active
operator $R(\omega)$, which rotates the molecule as a three-dimensional
object. In such case the magnetic field is 
always directed along the $z$-axis,
while the position of the molecule is specified by
three Euler angles $\omega=(\alpha,\beta,\gamma)$.
In the second case the position of the molecule is fixed but the
direction of the magnetic field is changed.  
In the latter case one has to know the matrix components
of the three projections of molecular orbital momentum. 
Below we have adopted the first approach which is
more familiar to us from our previous study of rotator functions 
\cite{Mic1,Nik}.
The advantage is that we are working only with the $z$-component
of orbital momentum. The details of the technique are given
in Appendices A and B.

Having calculated the matrix elements of $V_{mag}$ as a function
of the molecular rotation $\omega$,
we diagonalize the matrix 
${\cal H}=\sum_{a,b} V(\vec{r}_a,\vec{r}_b)+V_{mag}(a)$.
The degeneracies of molecular terms are lifted and the
magnetic moment of each sublevel $p$ is given by 
\begin{eqnarray}
  {\cal M}(p)=\langle p | {\cal M}_z | p \rangle ,
 \label{5.14} 
\end{eqnarray}
where $| p \rangle$ is the corresponding eigenvector.

For two electrons (or two $t_{1u}$ holes) we obtain 
${\cal M}_z=(\pm4.5$, $\pm2.5$, $\pm2$,
$\pm0.5$, 0) for $^3 T_{1g}$ (ground state), while
${\cal M}_z (^1 H_g)=(\pm5$, $\pm2.5$, 0).
In the $^1 A_g$ state 
the spin and the orbital momenta are quenched and ${\cal M}_z=0$.
For three electrons 
we have ${\cal M}_z (^4 A_u)=(\pm3$, $\pm1$) 
(the ground term); 
${\cal M}_z (^2 H_u)=(\pm6$, $\pm4$, $\pm3.5$, $\pm1.5$, $\pm1$)
 and ${\cal M}_z (^2 T_{1u})=(\pm3.5$, $\pm1.5$, $\pm1$).
We immediately conclude that the coupling scheme of orbital and spin 
momenta is different from the atomic case.
In order to clarify this issue we have studied the polarization of spin and 
orbital momenta separately. 
By excluding the spin momentum from Eq.\ (\ref{m.2})
we have found that 
${\cal L}_z (^3 T_{1g})=(\pm2.5(3)$, 0(3)),
and ${\cal L}_z (^1 H_g)=(\pm5$, $\pm2.5$, 0)
(numbers in parentheses stand for degeneracy).
In the molecular term $^4 A_u$
the orbital momentum is quenched, ${\cal L}_z=0(4)$, while 
${\cal L}_z (^2 H_u)=(\pm5(2)$, $\pm2.5(2)$, 0(2)), and
${\cal L}_z (^2 T_{1u})=(\pm2.5(2)$, 0(2)).
By excluding the orbital momentum from Eq.\ (\ref{m.2}) we have
obtained the results expected from the spin multiplicity index 
of molecular terms:
$2{\cal S}_z=(\pm2(3)$, 0(3)) for $^3 T_{1g}$
(spin triplet state)
and $2{\cal S}_z=0(5)$ for $^1 H_g$ (spin singlet).
For $t_{1u}^3$ we find $2{\cal S}_z(^4 A_u)=(\pm3$, $\pm1$), 
$2{\cal S}_z(^2 H_u) =\pm1(5)$, and $2{\cal S}_z(^2 T_{1u})=\pm1(3)$.

In Sec.~V we have already discussed the effect of mixing
between the $(t_{1u})^2$ and $(t_{1g})^2$ configurations
on the energy of the ground state of the C$_{60}^{2-}$ anion. 
The hybridization
affects also the magnetic moments of the $^3 T_{1g}$ ground state,
which are given by
\begin{eqnarray}
    {\cal M}=0,\; \pm g,\; \pm 2, \; \pm(2+g), \; \pm(4+g) .
 \label{5.14b} 
\end{eqnarray}
The magnetic moments of the unhybridized $^3 T_{1g}$ triplet of the
pristine $t_{1u}^2$ configuration correspond to $g=0.5$ 
(dashed line in Fig.~4).
The evolution of the $g$-factor as a function of $\triangle \epsilon_1$
is given in Fig.~4.
%
\begin{figure} 
\resizebox{0.43\textwidth}{!}
{ 
 \includegraphics{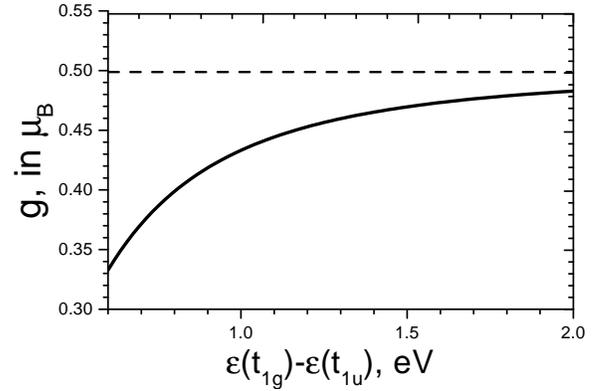} 
} 
\vspace{2mm}
\caption{
$g$ for the ground state triplet $^3 T_{1g}$
as a function of $\triangle \epsilon_1$, C$_{60}^{2-}$.
The corresponding magnetic moments are given by Eq.~(\ref{5.14b}). } 
\label{fig4} 
\end{figure} 
%

The results for the hole configurations $(h_u^+)^m$ are quoted
in Tables \ref{table_18}-\ref{table_21}. In the case of four
or five holes the number of molecular terms is too big (27) and we
give only magnetic moments for ten lowest states.
In general, magnetic moments are described by an expression
of the type (\ref{5.14b}) although in some cases
 two distinct values of $g$ are required.
%
\begin{table} 
\caption{ 
The magnetic moments ${\cal M}$ and the orbital momenta ${\cal L}$ of 
$(h_u^+)^2$, in $\mu_B$.
\label{table_18}     } 
\begin{ruledtabular}
 \begin{tabular}{c  | l l | l } 
   & ${\cal M}$ & $g$ & ${\cal L}$  \\
\hline 
$^3 T_{1g}$   & $0;\,\pm(1,3,4,5)\,g$  &  0.5    & $0,0,0,\,\pm (1,1,1)\,g$ \\
$^3 G_g$      & $\pm(1,1,3,3,5,5)\,g$  &  0.5    & $\pm(1,1,1,1,1,1)\,g$    \\
$^3 T_{2g}$   & $0,0,0,\,\pm(1,1,1)\,g$&  2      & 0(9)                     \\
$^1 G_g$      & $\pm(1,1)\,g$          &  0.8333 & ${\cal M}$  \\
$^1 H_g$      & $0;\, \pm(1,2)\,g$     &  0.0668 & ${\cal M}$   \\
$^1 H_g$      & $0;\, \pm(1,2)\,g$     &  0.2335 & ${\cal M}$   \\
$^1 A_g$      &  0                     &    0    & 0          
 \end{tabular} 
\end{ruledtabular}
\end{table} 
%
\begin{table} 
\caption{ 
Magnetic moments ${\cal M}$ of 
$(h_u^+)^3$, in $\mu_B$.
$^2 T_{1u}$ and ${}^2 T_{2u}$ states are degenerate.
\label{table_19}     } 
\begin{ruledtabular}
 \begin{tabular}{c  | l l} 
   &  ${\cal M}$ & $g$  \\
\hline 
$^4 T_{1u}$ & $\pm(1,2,3,5,6,7)\,g$             & 0.5 \\
$^4 G_u$    & $\pm(1,1,3,3,5,5,7,7)\,g$         & 0.5 \\
$^4 T_{2u}$ & $\pm(1,1,1,3,3,3)\,g$             & 1.0 \\
$^2 T_{2u}$ & $\pm(1,1,1)\,g$                   & 1.0 \\
$^2 H_u$    & $\pm(g_1,g_2,1,2-g_2,2-g_1)$      & 0.6529, 0.8265 \\
$^2 T_{1u}$ & $\pm(g,1,2-g)$                    & 0.25 \\
$^2 H_u$    & $\pm(g_1,g_2,1,2-g_2,2+g_1)$      & 0.2218, 0.3891 \\
$^2 G_u$    & $\pm(g,g,2-g,2-g)$                & 0.8983 \\
$^2 T_{1u}+{}^2 T_{2u}$ & $\pm(1,1,1,1,1,1)\,g$ & 1 \\
$^2 H_u$    & $\pm(g_1,g_2,1,2-g_2,2-g_1)$      & 0.9578, 0.9789 \\
$^2 G_u$    & $\pm(g,g,2-g,2-g)$                & 0.5650 \\
$^2 H_u$    & $\pm(0,1,2,3,4)\,g$               & 0.5 \\
 \end{tabular} 
\end{ruledtabular}
\end{table} 
%
\begin{table} 
\caption{ 
Magnetic moments ${\cal M}$ 
for ten lowest molecular terms of 
$(h_u^+)^4$, in $\mu_B$.
\label{table_20}     } 
\begin{ruledtabular}
 \begin{tabular}{c  | l l} 
   &  ${\cal M}$ & $g$  \\
\hline 
$^5 H_g$    & $0,\pm(1,2,2,3,4,5,6,6,7,8,9,10)\,g$        & 0.5          \\
$^3 G_g$    & $\pm(g,g,2-g,2-g,2+g,2+g)$                  & 0.310        \\
$^3 H_g$    & $0, \pm(g_1,g_2,2-g_2,2-g_1,2,2+g_1,2+g_2)$ & 0.046, 0.092 \\
$^1 G_g$    & $\pm(1,1)\,g$                               & 0.011        \\
$^1 A_g$    &  0                                          & 0            \\
$^3 T_{1g}$ &  $0,\pm(g,2-g,2,2+g)$                       & 0.052        \\
$^3 G_g$    &  $\pm(g,g,2-g,2-g,2+g,2+g)$                 & 0.023        \\
$^3 T_{2g}$ &  $0,0,0,\,\pm(1,1,1)\,g$                    & 2.0          \\
$^1 H_g$    &  $0,\, \pm(1,2)\,g$                         & 0.146        \\
$^1 T_{2g}$ &  0,0,0                                      & 0
 
 \end{tabular} 
\end{ruledtabular}
\end{table} 
%
\begin{table} 
\caption{ 
Magnetic moments ${\cal M}$ 
for ten lowest molecular terms of 
$(h_u^+)^5$, in $\mu_B$. 
${\cal M}(^4 H)$ stands for 
$\pm(g_1,g_2,1,2-g_2,2-g_1,2+g_1,2+g_2,$ $3,4-g_2,4-g_1)$.
\label{table_21}     } 
\begin{ruledtabular}
 \begin{tabular}{c  | l l} 
   &  ${\cal M}$ & $g$  \\
\hline 
$^6 A_u$    & $\pm(1,3,5)\,g$                     & 1.0 \\
$^4 H_u$    & ${\cal M}(^4 H)$                    & 0.533,0.766 \\
$^2 H_u$    & $\pm(g_1,g_2,1,2-g_2,2-g_1)$        & 0.873,0.936 \\ 
$^2 G_u$    & $\pm(g,g,2-g,2-g)$                  & 0.366       \\
$^4 H_u$    & ${\cal M}(^4 H)$                    & 0.866,0.933 \\
$^4 G_u$    & $\pm(g,g,2-g,2-g,2+g,2+g,4-g,4-g)$  & 0.167 \\
$^2 A_u$    & $\pm g$                             & 1.0   \\
$^2 G_u$    & $\pm(g,g,2-g,2-g)$                  & 0.978 \\
$^4 T_{2u}$ & $\pm(1,1,1,3,3,3)\,g$               & 1.0   \\
$^4 G_u$    & $\pm(g,g,2-g,2-g,2+g,2+g,4-g,4-g)$  & 0.5  

 \end{tabular} 
\end{ruledtabular}
\end{table} 

Interestingly, we have found that the calculated magnetic
moments are independent of the molecular orientation.
The conclusion holds for both $t_{1u}$ and $h_u$ shells
and we think that there must be a hidden group-theoretical reason
behind this. 
We consider the result as unexpected, because
the magnetic moment of a $\pi$ or $\delta$ MO of
diatomic molecules is anisotropic in respect to the
direction of the magnetic field.
From our previous study of C$_{60}^-$
in a cubic environment it also follows that the crystal field of C$_{60}^-$
exhibits strong dependence on its orientation \cite{Nik}.
In case of the C$_{60}$ molecule,
the orbital $t_{1u}$ (or $h_u$) space of the icosahedral symmetry 
is greatly reduced
in comparison with the 11 dimensional 
$l=5$ space of the rotation group $SO(3)$.
However, this is not accompanied by an anisotropic behavior of orbital momenta.

In order to understand this issue we have considered a simplified case of
one electron on the $t_{1u}$ molecular level.
Applying $\vec{H}$ in a direction
$\vec{n}=\vec{H}/H$, where $n_x=\sin\Omega\cos\phi$,
$n_y=\sin\Omega\sin\phi$ and $n_z=\cos\Omega$, we find
\begin{eqnarray}
 V_{mag}=\vec{H}\vec{L}=H\, L_{\vec{n}} .
 \label{5.15} 
\end{eqnarray}
Here the matrix $L_{\vec{n}}$ is given by
\begin{eqnarray}
   L_{\vec{n}}=n_x L_x + n_y L_y + n_z L_z .
 \label{5.16} 
\end{eqnarray}
By using Eqs.\ (\ref{1.2a}-c) for three $t_{1u}$ MOs,
after calculating the matrix elements, we arrive at
\begin{eqnarray}
 L_{\vec{n}}=\left[ \begin{array}{c c c}
 0 & -M_0 n_y\,i & M_0 n_x\,i \\
 M_0 n_y\,i & 0 & M_0 n_z\,i \\
 -M_0 n_x\,i & -M_0 n_z\,i & 0
 \end{array} \right] ,
 \label{5.17} 
\end{eqnarray}
where $M_0=2.5\, \mu_B$. The magnetic moments are obtained through the
diagonalization of $L_{\vec{n}}$. We find that ${\cal M}=0$ and
${\cal M}=\pm2.5\, \mu_B$ for any direction of $\vec{H}$.
The same conclusion is obtained for the case of one $h_u$-electron 
(or $h_u-$ hole). The matrix $L_{\vec{n}}$ then reads
\begin{eqnarray}
  L_{\vec{n}} \! = \!\! \left[ \begin{array}{c c c c c}
 0 & \frac{\sqrt{3}}{2}n_x\,i &  \frac{\sqrt{3}}{2}n_y\,i & 0 & 0 \\
 -\frac{\sqrt{3}}{2}n_x\,i & 0 & -\frac{1}{2}n_z\,i & \frac{1}{2}n_y\,i & -\frac{1}{2}n_x\,i \\
 -\frac{\sqrt{3}}{2}n_y\,i & \frac{1}{2}n_z\,i & 0 & \frac{1}{2}n_x\,i & \frac{1}{2}n_y\,i \\
  0 & -\frac{1}{2}n_y\,i & -\frac{1}{2}n_x\,i &  0 & -n_z\, i \\
  0 & \frac{1}{2}n_x\,i  & -\frac{1}{2}n_y\,i & n_z\, i & 0
 \end{array} \right] \! . \quad
 \label{5.18} 
\end{eqnarray}
The matrix has the same magnetic moments
(eigenvalues), which are 0, $\pm1/2$, $\pm 1$ (in $\mu_B$), 
for any direction of $\vec{H}$. 
The reasoning  given above
 is suggestive and we are looking
for a full-scale group-theoretical solution to this problem.

\section {Electron optical transitions } 
\label{sec:ot} 

In this section we consider only the electronic dipolar transitions
and the corresponding optical lines for C$_{60}^{n-}$.
The picture is not complete because there exist
electron-vibration interactions (``Herzberg-Teller" couplings) which
can alter the symmetry of the initial or the final state \cite{Dre}.
Here we omit the electron-vibration couplings and limit ourselves
to the electronic part of the problem.

The optically active transitions are associated with a nonzero 
expectation value of the electron dipolar operator $\vec{P}$.
Since the electric-dipole moment, 
\begin{eqnarray}
   \vec{P}=-e\sum_i \vec{r}_i ,
 \label{o.0} 
\end{eqnarray}
is an odd quantity in
respect to the inversion symmetry, it follows that $\vec{P}$ has 
no matrix components between states of the same parity.
Therefore, all spectral lines due to electric-dipole radiation
arise from transitions between states of opposite parity
(Laporte's rule) \cite{CS} and the following schemes
are relevant for the C$_{60}^{2-}$ 
and C$_{60}^{3-}$ molecular ions:
\begin{subequations}
\begin{eqnarray}
   & & (t_{1u})^2 \rightarrow t_{1u} t_{1g} ,   \label{o.1b}  \\
   & & (t_{1u})^3 \rightarrow (t_{1u})^2 t_{1g} .   \label{o.1c} 
\end{eqnarray}
\end{subequations}
These configurations have been considered already in previous 
sections, and now we can proceed to
calculations of optical transitions.

In atoms there are several additional selection rules which greatly
facilitate line assignments. These rules are not developed for
the icosahedral symmetry and in the following
we have to rely on numerical analysis.
The total intensity for the line from
level $A$ to level $B$ is given by~\cite{CS}
\begin{eqnarray}
   I(A,B)=N(a)\, h\nu\, \frac{64 \pi^4 \sigma^3}{3h}\, S(A,B),
 \label{o.2} 
\end{eqnarray}
where $N(a)$ is the number of C$_{60}^{n-}$ molecules in state $a$,
$\nu=(E_A-E_B)/h$ is the frequency and $\sigma=h\nu/c$ is the wave number.
Finally, $S(A,B)$ is the line strength which is found as
\begin{eqnarray}
 S(A,B)=\sum_{ab} |\langle a |\vec{P}| b \rangle |^2 .
 \label{o.3} 
\end{eqnarray}
The line strength is a very convenient quantity and 
in the following we 
calculate $S(A,B)$ for the transitions (\ref{o.1b},b).

The matrix elements of the dipole operator  for the case (\ref{o.1b}) 
read
\begin{eqnarray}
 \langle a | \vec{P} |b \rangle = \sum_{IJ'} \langle a | I \rangle
 \langle I |\vec{P}| J' \rangle \langle J' | b \rangle ,
 \label{o.4} 
\end{eqnarray}
where $| I \rangle$ and $| J' \rangle$ are the basis states of
$(t_{1u})^2$ and $(t_{1u} t_{1g})$, respectively,
while $\langle a | I \rangle$ and $\langle b | J' \rangle$ are
the eigenvectors corresponding to levels $a$ and $b$.
The dipole moment (\ref{o.0}) is a one-electron operator, Sec.~III.
Its matrix elements are given by
\begin{equation}
 \langle I | \vec{P} |J' \rangle = 
 \delta(i_{u1},j_u)\, \langle i_{u2} | \vec{P} | j_g \rangle -
 \langle i_{u1} |\vec{P}| j_g \rangle\, \delta(i_{u2},j_u) .
 \nonumber 
\end{equation}
We recall that $\langle I |=\langle i_{u1},i_{u2} |$, 
where $i_{u1}$ and $i_{u2}$ ($i_{u1}>i_{u2}$)
are indices referring to six $t_{1u}$ states, and 
$|J' \rangle=|j_u,j_g \rangle$, Eq.\ (\ref{2.1}).
From parity consideration it follows that the 
nonzero matrix elements are
of the type $\langle i_u |\vec{P}|j_g \rangle$. 
In order to
calculate them, we first rewrite $\vec{P}$ in the following form: 
\begin{subequations}
\begin{eqnarray}
  & & P_x= \sqrt{\frac{4\pi}{3}}\,r\, Y_1^{1,c}(\hat{r}),  \label{o.6a} \\ 
  & & P_y= \sqrt{\frac{4\pi}{3}}\,r\, Y_1^{1,s}(\hat{r}),  \label{o.6b} \\ 
  & & P_z= \sqrt{\frac{4\pi}{3}}\,r\, Y_1^0(\hat{r}) .  \label{o.6c}  
\end{eqnarray}
\end{subequations}
Here $Y_1^{\tau}$ are real spherical harmonics \cite{Bra}, 
and as before $r=|\vec{r}\,|$, while $\hat{r}$ stands for the polar
angles $(\Omega,\phi)$. Then we find that
the one-electron matrix elements of $\vec{P}$ are
\begin{eqnarray}
 \langle i_u |P_k| j_g \rangle = {\cal V}\, c_{1,{\tau (k)}}(i_u,j_g),  
  \label{o.7} 
\end{eqnarray}
where $\tau=(1,c)$, $(1,s)$ or $0$ for $k=x,y$ and $z$, respectively. 
The quantities $c_{l=1,\tau}(i_u,j_g)$ are given by Eq.~(\ref{2.11})
for $l=1$. In fact, these coefficients have been also 
used for the calculation of the $(t_{1u} t_{1g})$ and $(t_{1u})^2 t_{1g}$
configurations in Sec.\ IV and V.
Finally, the radial part of Eq.\ (\ref{o.7}) reads
\begin{eqnarray}
 {\cal V} = \sqrt{\frac{4\pi}{3}}\, 
 \int dr\,r^3\,   {\cal R}_{t1u}(r)\,{\cal R}_{t1g}(r) .
  \label{o.8} 
\end{eqnarray}
Since we have already computed the eigenvalues and eigenvectors
of $(t_{1u})^2$ and $(t_{1u} t_{1g})$ in Sec.~V,
we now can calculate the line strengths using
equations (\ref{o.3})-(\ref{o.8}).
The results are quoted in Table~\ref{table_15}.
%
\begin{table} 
\caption{ 
 Energies $E_{ab}=\triangle \epsilon_1+\epsilon_{ab}$ (in eV) 
 and line strengths (in ${\cal V}$) of the transitions 
 $[(t_{1u})^2;\, a] \rightarrow [(t_{1u} t_{1g});\,b]$,
 $a=1-3$, $b=1-6$, calculated with the model III.
 Only transitions with $S(a,b) \neq 0$ are given.
\label{table_15}     } 
\begin{ruledtabular}
 \begin{tabular}{c | c c | c c | c c } 
    & \multicolumn{2}{c|}{$(t_{1u})^2;$  $a={}^3 T_{1g}$} &
      \multicolumn{2}{c|}{$^1 H_g$} &
      \multicolumn{2}{c}{$^1 A_g$} \\ 
\hline 
 $(t_{1u} t_{1g});\,b$    & $\epsilon_{ab}$ &  $S$ & 
                        $\epsilon_{ab}$ &  $S$ & 
                        $\epsilon_{ab}$ &  $S$ \\
\hline
$^1 A_u$   &        &       &       &       &       &      \\
$^3 H_u$   & -0.350 & 0.482 &       &       &       &      \\
$^3 T_{1u}$& -0.267 & 0.289 &       &       &       &      \\
$^1 H_u$   &        &       & 0.392 & 0.482 &       &      \\
$^3 A_u$   &  0.624 & 0.386 &       &       &       &      \\
$^1 T_{1u}$&        &       & 0.609 & 0.161 & 0.323 & 0.129 

 \end{tabular} 
\end{ruledtabular}
\end{table} 

Similarly, one can treat the optical transitions (\ref{o.1c}) for C$_{60}^{3-}$.
Now we consider the matrix elements of $\vec{P}$
between three-electron basis states 
$\langle I\,(t_{1u})^3| =\langle i_1,i_2,i_3 |$
and $| J\,[(t_{1u})^2 t_{1g}] \rangle=|j_1,j_2,j_g \rangle$,
and obtain
\begin{eqnarray}
 & & \langle I | \vec{P} |J \rangle =  \nonumber \\  
 & & \langle i_1 | \vec{P} | j_g \rangle\, \delta(i_2,j_1)\, \delta(i_3,j_2) -
 \langle i_1 |\vec{P}| j_g \rangle\, \delta(i_2,j_2)\,  \delta(i_3,j_1) 
 \nonumber \\ 
 &+& \langle i_2 | \vec{P} | j_g \rangle\, \delta(i_1,j_2)\, \delta(i_3,j_1) -
 \langle i_2 |\vec{P}| j_g \rangle\, \delta(i_1,j_1)\,  \delta(i_3,j_2) 
 \nonumber \\ 
 &+& \langle i_3 | \vec{P} | j_g \rangle\, \delta(i_1,j_1)\, \delta(i_2,j_2) -
 \langle i_3 |\vec{P}| j_g \rangle\, \delta(i_1,j_2)\,  \delta(i_2,j_1)  .
 \nonumber 
\end{eqnarray}
Here again, the one-electron matrix elements $\langle i_u |\vec{P}|j_g \rangle$
are specified by Eq.\ (\ref{o.7}).
The resulting line strengths are quoted in Table~\ref{table_16}.
%
\begin{table} 
\caption{ 
 Energies $E_{ab}=\triangle \epsilon_1+\epsilon_{ab}$ (in eV) and 
 line strengths (in ${\cal V}$) of the transitions 
 $\{ (t_{1u})^3;\, a \} \rightarrow \{ (t_{1u}^2 t_{1g} ;\,b \}$,
 $a=1-3$, $b=1-10$, calculated with the model III.
 Only the transitions with $S(a,b) \neq 0$ are given.
\label{table_16}   } 
\begin{ruledtabular}
 \begin{tabular}{c | c c | c c | c c } 
    & \multicolumn{2}{c|}{$(t_{1u})^3;$  $a={}^4 A_u$} &
      \multicolumn{2}{c|}{$^2 H_u$} &
      \multicolumn{2}{c}{$^2 T_{1u}$} \\ 
\hline 
 $[(t_{1u})^2 t_{1g}];\,b$    & $\epsilon_{ab}$ &  $S$ & 
                        $\epsilon_{ab}$ &  $S$ & 
                        $\epsilon_{ab}$ &  $S$ \\
\hline
$^4 H_g$    &        &       &        &       &        &          \\
$^2 T_{1g}$ &        &       & -0.913 & 0.002 & -1.103 & $S<10^{-3}$ \\
$^1 G_g$    &        &       &        &       &        &          \\
$^2 G_u+{}^2 T_{2u}$ &        &       & -0.330 & 0.450 &        &        \\
$^4 T_{1g}$ & -0.010 & 0.771 &        &       &        &       \\
$^2 H_g$    &        &       & -0.158 & 0.120 & -0.349 & 0.205 \\
$^2 T_{1g}$ &        &       &  0.058 & 0.072 & -0.132 & 0.120 \\
$^2 H_g$    &        &       &  0.508 & 0.683 &  0.318 & 0.277 \\
$^2 A_g$    &        &       &        &       &  0.589 & 0.193 \\
$^2 T_{1g}$ &        &       &  0.780 & 0.602 &  0.920 & 0.362 

 \end{tabular} 
\end{ruledtabular}
\end{table} 

It follows from Table \ref{table_15} that for C$_{60}^{2-}$
there are three lines from the ground state $^3 T_{1g}$, 
\begin{subequations}
\begin{eqnarray}
& &E({}^3 T_{1g} \rightarrow {}^3 \!H_u)=\triangle \epsilon_1-0.350 \mbox{ eV}, \\
& &E({}^3 T_{1g} \rightarrow {}^3 T_{1u})=\triangle \epsilon_1-0.267 \mbox{ eV}, \\
& &E({}^3 T_{1g} \rightarrow {}^3 \!A_{u})=\triangle \epsilon_1+0.624 \mbox{ eV}.
  \label{o.10} 
\end{eqnarray}
\end{subequations}
With $\triangle \epsilon_1=1.153$ eV, Eq.\ (\ref{2.18a}), we obtain
$E({}^3 T_{1g} \rightarrow {}^3 H_u)=0.803$ eV,
$E({}^3 T_{1g} \rightarrow {}^3 T_{1u})=0.886$ eV
and $E({}^3 T_{1g} \rightarrow {}^3 A_{u})=1.777$ eV.
These values should be compared with two dominant bands at
1.305 eV (950 nm) and 1.494 eV (830 nm) observed by the near-infrared 
experiments in solutions \cite{Reed}.
We conclude that first two transitions can be tentatively ascribed to 
the experimental bands if $\triangle \epsilon_1$ is taken to be larger, 
$\triangle \epsilon_1 \sim 1.7$ eV. Here it is worth to notice
that in our approach $\triangle \epsilon_1$ in Eq.\ (\ref{2.18a})
remains a phenomenological quantity which is not immediately connected
with the term splittings. 
In Ref.\ \onlinecite{Neg} the authors have obtained that the ground state
of C$_{60}^{2-}$ is the $^1 A_g$ singlet. 
We have shown in Sec.\ V that this is possible if 
$\triangle \epsilon_1$ is small, see Fig.~3.
However, if $^1 A_g$ is the ground state, then there is only
one optical transition ($^1 A_g \rightarrow {}^1 T_{1u}$ at
$\triangle \epsilon_1 +0.323$ eV, Table \ref{table_15}) and
comparison with the experiment becomes even more
problematic.
We conclude that our calculations are basically in agreement with
the experiment for C$_{60}^{2-}$, although also a third band is expected.
The position of the third band however can change due to the effect of
configuration mixing discussed in Sec.~V.

The situation is less clear for the C$_{60}^{3-}$ molecular ion.
Both our calculations and those of Ref.\ \onlinecite{Neg} predict that
the ground state is the $^4 A_u$ level.
Then from Table \ref{table_16} we find that the only optical transition
allowed by the selection rules is $^4 A_u \rightarrow {}^4 T_{1g}$.
This is at variance with the experimental consensus for three
dominant bands at 1350, 960 and 770 nm \cite{Reed}.
In principle, the theoretical line $^4 A_u \rightarrow {}^4 T_{1g}$
can be split by the crystal field and Jahn-Teller distortions,
but the magnitude of the splitting ($\sim 0.3$ eV) seems 
excessive. The other possibility is if some transitions
become allowed through the ``Herzberg-Teller" (electron-vibration)
mechanism \cite{Dre}.
Further experimental and theoretical investigations 
are needed to clarify the issue.

\section {Discussion and conclusions} 
\label{sec:dc} 

We have presented an efficient configuration interaction method 
for many electron (hole) molecular terms of the C$_{60}^{m\pm}$
molecular ion.
The method is based on the multipole expansion of the Coulomb
interactions and takes into account the molecular symmetry.
Although there are some similarities with approaches used
for treating many electron effects in atomic calculations \cite{Lind},
the technique is novel and original. Crystal electric field
effects \cite{Nik} and the spin-orbit coupling can be easily incorporated
in the calculation. The technique can be used for other electron
systems.

We have applied the method for studying molecular terms of 
electron and hole configurations
of C$_{60}^{n-/m+}$ ($n=2-4$, $m=2-5$), and for
excitonic configurations $(h_u^+ t_{1u}^-)$
and $(h_u^+ t_{1g}^-)$ of the neutral molecule.
In most of the cases the ground state is found to obey Hund's rules.

Our calculations of the molecular term structure $(t_{1u})^2$ for C$_{60}^{2-}$
differs from the previous result of Negri {\it et al.}, Ref.~\onlinecite{Neg}.
They have reported that the ground term of C$_{60}^{2-}$ is the $^1 A_g$
singlet, while we have found that for realistic parameters
($\triangle \epsilon_1=1.15$ eV \cite{Reed,Kato}) 
it is the $^3 T_{1g}$ triplet, in accordance with Hund's rules. 
We have shown (Fig.~3 and Sec.~V) 
that the reason for this is that 
in Ref.\ \onlinecite{Neg} the one-electron energy difference 
between $t_{1g}$ and $t_{1u}$ states 
is too small,  $\triangle \epsilon_1=$0.64 eV.

Our results for the $(h_u^+)^2$ hole configuration indicate
that there are three very close ($\sim0.03$ eV) low lying molecular 
triplets of $^3 T_{1g}$, $^3 G_g$ and $^3 T_{2g}$ symmetry. 
The lowest molecular terms for $(h_u^+)^3$ belong to the
$^4 T_{1u}$, $^4 G_u$ and $^4 T_{2u}$ symmetry.
The number of states in a small energy interval
$\triangle \varepsilon \sim 0.03$ eV near the ground state
is 30, 40, 25, 6 for hole configurations $(h_u^+)^m$ 
with $m=$2, 3, 4, 5, respectively.
This suggests that the configuration of $m=3$ holes is
most susceptible for Jahn-Teller distortions of the C$_{60}$
molecule and possibly for hole-phonon coupling which
causes superconductivity \cite{Bat1}.

We have calculated the magnetic moments of the $(t_{1u})^n$ and
$(h_u^+)^m$ configurations, Sec.~VI.
The coupling of spin and orbital momenta differs from the Land\'{e}
$g-$factor scheme of atoms.
The magnetic moments do not depend on the orientation of the
molecule with respect to an external magnetic field.
The latter statement was demonstrated explicitly for
the case of one $t_{1u}$ electron and one $h_u$ hole.
We consider this as a 
group-theoretical puzzle of the icosahedral symmetry.
We have also found new ``accidental" degeneracy
between the $^2 G_g$ and $^2 T_{2g}$ states of the $(t_{1u})^2 t_{1g}$
and $(t_{1g})^2 t_{1u}$  configurations 
(Tables \ref{table9} and \ref{table_12}.)

Finally, we have studied optical absorption associated with
electron dipolar transitions $(t_{1u})^2 \rightarrow (t_{1u} t_{1g})$ and
$(t_{1u})^3 \rightarrow (t_{1u})^2 t_{1g}$. 
For C$_{60}^{2-}$ we have found that two lines
($^3 T_{1g} \rightarrow {}^3 H_u$ and $^3 T_{1g} \rightarrow {}^3 T_{1u}$)
can be tentatively ascribed to the two near-infrared 
dominant bands at 950 and 830 nm. However, in addition 
a third band ($^3 T_{1g} \rightarrow {}^3 A_{u}$)
is expected from the calculation.
For C$_{60}^{3-}$ with $^4 A_u$ as the ground state, 
we have found that only one electron dipolar
transition, $^4 A_u \rightarrow {}^4 T_{1g}$, is allowed. 
It seems that a better understanding of optical transitions 
requires a study of the Herzberg-Teller (electron-vibration) effect
which is beyond the scope of the present work.
We suggest to perform optical experiments for C$_{60}^{n-}$
and C$_{60}^{m+}$ in the gas phase to obtain more precise and
full data on the optical lines which can shed light on the problem
of electronic intra-molecular correlations.

\acknowledgments 

We thank M. Lueders, N. Manini, E. Tosatti, and F. Negri for interesting
discussions, P. Launois, A. Ceulemans and P.N. Dyachkov for 
informing us about useful references.
This work has been financially supported by
the Fonds voor Wetenschappelijk Onderzoek, Vlaanderen.

\appendix

\section{$L_z$ for a rotated molecule}

If the molecule is rotated away, 
then the MOs of $t_{1u}$ and $h_u$ symmetry
are given by Eqs.\  (\ref{1.2a}-c) and (\ref{1.1a}-e) in the
coordinate system $(x',y',z')$ attached to the molecule.
The rotated functions can be expanded in terms of
real spherical harmonics (RSH) defined in the fixed set of axes $(x,y,z)$.
For example, a rotation $R$ defined by three Euler angles
$\omega=(\alpha,\beta,\gamma)$ transforms
$\psi_1(t_{1u})$ to
\begin{eqnarray}
 & & \psi'_1(t_{1u})=R(\omega)\, \psi_1(t_{1u})  \nonumber \\
 & &  =\frac{6}{\sqrt{50}}\, R(\omega) Y_5^0
 +\sqrt{\frac{7}{25}} R(\omega) Y_5^{5,c} .
  \label{5.9} 
\end{eqnarray}
Here $R(\omega) Y_5^0$ and $R(\omega) Y_5^{5,c}$ defines
the rotations of $Y_5^0$ and $Y_5^{5,c}$,
respectively. 
The rotated functions can be found from 
Eqs.\ (\ref{a.5}), and (\ref{a.9}), quoted in the Appendix B.
Analogously, we proceed with the other angular functions of $t_{1u}$
and $h_u$ symmetry. In general,
\begin{eqnarray}
  R(\omega)\,Y_l^{\tau}=\sum_{\tau'} Y_l^{\tau'}\,U^l_{\tau' \tau}(\omega) .
  \label{5.10} 
\end{eqnarray}
Here $\tau,\tau'=(m,c)$, $(m,s)$ or 0, and the rotator functions
(matrices) $U^l_{\tau' \tau}(\omega)$ are specified in Appendix B,
Eqs.\ (\ref{a.5}), (\ref{a.7}), and (\ref{a.9}).
Now we are ready to calculate $L_z \psi'_k(t_{1u})$ ($k=1-3$) and
$L_z \psi'_k(h_u)$ ($k=1-5$). By means of Eq.\ (\ref{5.6})
we obtain
\begin{eqnarray}
 \lefteqn{ \left. \langle k | L_z | k' \rangle \right|_{\omega} }  \nonumber \\
 &=& i \sum_{m>0} 
 m\,(g_{(m,s)k}\, g_{(m,c)k'}-g_{(m,c)k}\, g_{(m,s)k'}) , \quad
 \label{5.11} 
\end{eqnarray}
where the functions $g_{\tau,k}$ depend on $\omega$,
\begin{subequations}
\begin{eqnarray}
  g_{(m,c)k}(\omega)=\alpha(\tau_1)\,U^l_{(m,c)\tau_1}(\omega)
 +\alpha(\tau_2) U^l_{(m,c)\tau_2}(\omega)  , \nonumber \\ 
  g_{(m,s)k}(\omega)=\alpha(\tau_1)\,U^l_{(m,s)\tau_1}(\omega)
 +\alpha(\tau_2) U^l_{(m,s)\tau_2}(\omega)  . \nonumber  
\end{eqnarray}
\end{subequations}
Here $\alpha(\tau)$ stands for the coefficients of expansion
of MOs of $t_{1u}$ and $h_u$ symmetry in terms of RSH,
\begin{eqnarray}
  \psi_k(\Omega)=\alpha(\tau_1)\, Y_l^{\tau_1}(\Omega)
  +\alpha(\tau_2)\, Y_l^{\tau_2}(\Omega) ,
  \label{5.10a} 
\end{eqnarray}
see Eqs.\ (\ref{1.2a}-c) and (\ref{1.1a}-e).
For example, for the first MO ($k=1$) of 
$t_{1u}$ symmetry we have
$\tau_1=0$, $\alpha(\tau_1)=6/\sqrt{50}$ and $\tau_2=(5,c)$,
$\alpha(\tau_2)=\sqrt{7/25}$, and etc. 
The indices $k,k'$ in (\ref{5.11}) belong to the same molecular shell
($t_{1u}$ or $h_u$), otherwise $\langle k | L_z | k' \rangle|_{\omega}=0$.
From Eq.\ (\ref{5.11}) we also conclude that
\begin{subequations}
\begin{eqnarray}
 & & \left. \langle k | L_z | k \rangle\right|_{\omega}=0 ,  
            \label{5.12'a}  \\
 & & \left. \langle k | L_z | k' \rangle \right|_{\omega}
 =\left. \langle k' | L_z | k \rangle\right|_{\omega}^* .
 \label{5.12'b} 
\end{eqnarray}
\end{subequations}
The former condition is a consequence of working with real 
spherical harmonics, the latter ensures the hermiticity of $L_z$.

\section{Rotation of real spherical harmonics}

An active rotation $R$ is specified by its Euler angles 
$\omega=(\alpha,\beta,\gamma)$ \cite{Bra}. 
It transforms a complex spherical harmonic $Y_l^m$ to ${Y'}_l^m$, where
\begin{subequations}
\begin{eqnarray}
   {Y'}_l^m=R(\omega)\, Y_l^m =\sum_n Y_l^n D_{nm}^l(\omega) .
 \label{a.1a} 
\end{eqnarray}
For ${Y'}_l^{-m}$ we have
\begin{eqnarray}
   {Y'}_l^{-m}=R(\omega)\, Y_l^{-m} =\sum_n Y_l^n D_{n-m}^l(\omega) .
 \label{a.1b} 
\end{eqnarray}
\end{subequations}
Here $D_{nm}^l$ stands for the Wigner functions given by
\begin{eqnarray}
    D_{nm}^l(\alpha,\beta,\gamma)=C_{nm}\, 
    e^{-in\gamma}\,d^l(\beta)_{nm}\, e^{-im\alpha} .
 \label{a.2} 
\end{eqnarray}
$d^l(\beta)_{nm}$ is a reduced matrix element which is a real quantity,
and $C_{nm}=\pm 1$ depending on $n,m$ (see Eqs.\ (2.1.6) and (2.1.5)
of Refs.\ \onlinecite{Bra}). From Eq.\ (\ref{a.2}) and the properties
\begin{subequations}
\begin{eqnarray}
  & & d(\beta)_{-n,-m}=(-1)^{n+m}d(\beta)_{nm} ,    \label{a.3a} \\ 
  & & C_{-n,-m}=(-1)^{n+m}C_{nm}=(-1)^{n+m}C_{nm}^* , \quad   \label{a.3b}  
\end{eqnarray}
\end{subequations}
we find that
\begin{eqnarray}
    D_{-n-m}^l(\omega)=D_{nm}^l(\omega)^* .
 \label{a.4} 
\end{eqnarray}
We then combine (\ref{a.1a}) with (\ref{a.1b}) and use Eq.\ (\ref{a.4})
for deriving 
the transformation law of real spherical harmonics. After some
algebra we find
\begin{eqnarray}
  & & R\, Y_l^{m,c} = Y_l^0\, U^l_{0;(m,c)}   \nonumber \\
  & & +\sum_{n>0} \left( Y_l^{n,c}\,U^l_{(n,c);(m,c)}
   +Y_l^{n,s}\,U^l_{(n,s);(m,c)} \right) ,   \quad  \label{a.5}  
\end{eqnarray}
where $U^l_{0;(m,c)}=\sqrt{2}\,Re D^l_{0m}$, and
\begin{subequations}
\begin{eqnarray}
  & & U^l_{(n,c);(m,c)}=Re(D^l_{nm}+D^l_{n-m}) ,     \label{a.6a} \\
  & & U^l_{(n,s);(m,c)}=-Im(D^l_{nm}+D^l_{n-m}) .    \label{a.6b}  
\end{eqnarray}
\end{subequations}
In Eqs.\ (\ref{a.5})-(\ref{a.6b}) and below for clarity we drop
the argument $\omega$.
Analogously, rotating $Y_l^{m,s}$ we obtain
\begin{eqnarray}
  & & R\, Y_l^{m,s} = Y_l^0\, U^l_{0;(m,s)} \nonumber \\
  & & +\sum_{n>0} \left( Y_l^{n,c}\,U^l_{(n,c);(m,s)}
   +Y_l^{n,s}\,U^l_{(n,s);(m,s)} \right) , \quad   \label{a.7}  
\end{eqnarray}
where $U^l_{0;(m,s)}=\sqrt{2}\,Im D^l_{0m}$, and
\begin{subequations}
\begin{eqnarray}
  & & U^l_{(n,c);(m,s)}=Im(D^l_{nm}-D^l_{n-m}) ,     \label{a.8a} \\
  & & U^l_{(n,s);(m,s)}=Re(D^l_{nm}-D^l_{n-m}) .    \label{a.8b}  
\end{eqnarray}
\end{subequations}
Finally, the rotation of $Y_l^0$ yields
\begin{equation}
  R\, Y_l^0 = Y_l^0\, U^l_{0;0} 
  +\sum_{n>0} \left( Y_l^{n,c}\,U^l_{(n,c);0}
   +Y_l^{n,s}\,U^l_{(n,s);0} \right) , 
 \label{a.9}  
\end{equation}
where $U^l_{0;0}= D^l_{00}$, 
$U^l_{(n,c);0}=\sqrt{2}\, Re D^l_{n0}$
and $U^l_{(n,s);0}=-\sqrt{2}\, Im D^l_{n0}$.



\end{document}